\documentclass{aastex}

\usepackage{epsfig,amsfonts,amsmath,amssymb,graphicx,morefloats,appendix,tensor,gensymb}
\usepackage{color}
\usepackage{hyperref}
\newcommand\vsini{\ifmmode{v\sin{i_\star}}\else $v\sin{i_\star}$\fi}
\newcommand\sini{\ifmmode{\sin{i_\star}}\else $\sin{i_\star}$\fi}

\newcommand\mysim{\mathord{\sim}}

\newcommand{\rffigl}[1]{Figure~\ref{fig:#1}}

\newcommand{\rfsecl}[1]{\mbox{Section \ref{sec:#1}}}

\newcommand{\rftabl}[1]{Table~\ref{tab:#1}}

\begin{document}

\title{Evidence for Misalignment Between Debris Disks and Their Host Stars}
\shorttitle{Misalignment Between Debris Disks and Their Stars}

\author[0000-0002-6903-9080]{Spencer A. Hurt}
\affiliation{Department of Physics and Astronomy, University of Wyoming, Laramie, WY 82071, USA}
\affiliation{Department of Astrophysical and Planetary Sciences, University of Colorado, Boulder, CO 80309, USA}

\author[0000-0001-7891-8143]{Meredith A. MacGregor}
\affiliation{Department of Astrophysical and Planetary Sciences, University of Colorado, Boulder, CO 80309, USA}

\shortauthors{Hurt $\&$ MacGregor}

\begin{abstract}
We place lower limits on the obliquities between debris disks and their host stars for 31 systems by comparing their disk and stellar inclinations. While previous studies did not find evidence for misalignment, we identify 6 systems with minimum obliquities falling between $\mysim30\degree-60\degree$, indicating that debris disks can be significantly misaligned with their stars. These high-obliquity systems span a wide range of stellar parameters with spectral types K through A. Previous works have argued that stars with masses below 1.2 $M_\odot$ (spectral types of $\mysim$F6) have magnetic fields strong enough to realign their rotation axes with the surrounding disk via magnetic warping; given that we observe high obliquities for relatively low-mass stars, magnetic warping alone is likely not responsible for the observed misalignment. Yet, chaotic accretion is expected to result in misalignments of $\mysim20\degree$ at most and cannot explain the larger obliquities found in this work. While it remains unclear how primordial misalignment might occur and what role it plays in determining the spin-orbit alignment of planets, future work expanding this sample is critical towards understanding the mechanisms that shape these high-obliquity systems.
\end{abstract}

\section{Introduction}
\label{sec:introduction}

The Sun's equatorial plane is well-aligned with the ecliptic, having an obliquity of $7.155\pm0.002\degree$ \citep{Beck:2005}. Most of the major solar system bodies move in nearly the same plane, suggesting that the planets formed from a protoplanetary disk that was rotating in the same direction as the early Sun. It has commonly been thought that other planetary systems form similarly and that exoplanet orbital axes should be closely aligned with their stars' spin axes. However, observational techniques such as the Rossiter-McLaughlin effect \citep{Queloz:2000, Shporer:2011, Triaud:2018}, Doppler shadows \citep{Albrecht:2007, Zhou:2016}, and gravity darkened transits \citep{Barnes:2009, Ahlers:2020} have measured large spin-orbit angles for many extra-solar systems \citep{Albrecht:2022}.

Possible mechanisms responsible for spin-orbit misalignment generally fall into three categories: primordial misalignment, post-formation misalignment, and changes in the stellar spin axis that are independent of planet formation. The first, primordial misalignment, suggests that a protoplanetary disk is misaligned with its star's rotation axis and that planets with large spin-orbit angles form in situ. Processes that could misalign the disk include chaotic accretion (where the late arrival of material from the molecular cloud warps or tilts the disk; \citealt{Bate:2010, Thies:2011, Fielding:2015, Bate:2018, Takaishi:2020}), magnetic warping (when the Lorentz force between a young star and ionized inner disk magnifies any initial misalignments; \citealt{Lai:2011, Foucart:2011}), and secular processes involving an inclined stellar or planetary companion \citep{Borderies:1984, Lubow:2000, Batygin:2012, Matsakos:2017}. Post-formation misalignment implies that after formation, gravitational interactions alter a planet's orbit. This could occur via planet-planet scattering \citep{Malmberg:2011, Beauge:2012} or secular processes like Kozai-Lidov cycles \citep{Naoz:2016} or disk-driven resonance \citep{Petrovich:2020}. Both primordial and post-formation misalignment could also occur via stellar clustering, which appears to have a strong influence on the architecture of planetary systems \citep{Tristan:2019, Winter:2020, Rodet:2022} and may be commonplace \citep{Yep:2022}. Thirdly, it has been proposed that stars with convective cores and radiative envelopes can reorient themselves without an external torque due to internal gravity waves generated at the radiative-convective boundary \citep{Rogers:2012, Rogers:2013}.

Hot Jupiters---massive planets on very short orbits---frequently appear misaligned with hot, rapidly-rotating stars that generally fall above the Kraft break \citep[$\mysim6200$~K;][]{Kraft:1967} while low-mass planets appear misaligned with both cool and hot stars \citep{Winn:2010, Schlaufman:2010, Albrecht:2022}. It has been suggested that hot Jupiters first enter high-obliquity orbits regardless of their host stars' properties; however, tidal interactions between the massive, close-orbiting planets and the relatively thick convective envelopes found in stars below the Kraft break realign the stellar spin axes with the hot Jupiters' orbits. The mechanisms responsible for spin-orbit misalignment may help reveal how these exotic planets form. While formation in situ through core accretion may be possible \citep{Batygin:2016}, it would be challenging for enough material to accumulate and develop into a planet that close to a star. Instead, if a massive planet formed far from its star, it could move to a short orbit via disk-driven migration or high-eccentricity tidal migration \citep{Dawson:2018}. If the hot Jupiter were primordially misaligned, this would indicate disk-driven migration, whereas post-formation misalignment could result from high-eccentricity tidal migration.

Constraints on which mechanisms actually contribute to spin-orbit misalignment can be placed using the observed distribution of obliquities and trends across system parameters. Additional constraints on primordial misalignment can be placed using observations of circumstellar disks and their stars. \cite{Watson:2011} first compared stellar inclinations to disk inclinations for 8 systems with spatially resolved debris disks, while \cite{Greaves:2014} later did the same for 10 systems imaged by the \textit{Herschel} satellite. Neither found evidence for misalignment, but both had limited samples and predate many spatially resolved images of disks taken by the Atacama Large Millimeter/submillimeter Array (ALMA), Hubble Space Telescope (HST), and Gemini Planet Imager (GPI) that can robustly measure disk inclinations. \cite{Davies:2019} compared inclinations for resolved disks (mostly protoplanetary) in the $\rho$ Ophiuchus and Upper Scorpius star forming regions, finding that a third of systems are potentially misaligned. \cite{Davies:2019} used these contrasting results to raise the additional question of whether or not debris disks preserve the preceding protoplanetary disks' geometry and if star-disk-planet interactions or the formation of a debris disk can change the star-disk obliquity.

In this work, we study the star-disk alignment for an expanded sample of 31 resolved debris disks. In \rfsecl{methods}, we outline our methods, including the sample selection and measurements made. We then discuss our results in \rfsecl{results}. Finally, in \rfsecl{conclusions}, we conclude our findings.

\section{Methods}
\label{sec:methods}

We assembled a list of spatially resolved debris disks from the literature, excluding circumbinary and circumtriple disks to simplify our analysis. We then identified systems with published stellar inclinations ($i_s$) or the data necessary to measure the inclination available, leaving a sample of 31 targets that can be found in \rftabl{sample}.

Effective temperatures ($T_\mathrm{eff}$), masses ($M$), and radii ($R$) were taken from the \textit{TESS} Input Catalog (TIC; \citealt{Stassun:2018, Paegert:2021}) for most stars in our sample. Given the majority resolved debris disks are located around nearby, bright stars, the TIC adopted most of these parameters from large spectroscopic catalogs, avoiding the challenges of color-temperature relationships discussed in \cite{Stassun:2018}, particularly for the coolest stars ($T_\mathrm{eff}<3800$ K). Parallaxes are known for all objects in our sample, providing precise measurements of radius in the TIC. 4 targets (AU Mic, Vega, $\beta$ Leonis, and $\beta$ Pictoris) that were either missing measurements or were reported without uncertainties were supplemented using values found elsewhere in the literature. The number of confirmed planets in each system was additionally determined by searching the NASA Exoplanet Archive Confirmed Planets Table \citep{https://doi.org/10.26133/nea1}. 8 out of the 31 systems have at least one confirmed planet. 

To determine the projected rotational velocities ($v\sin i$) of our sample, we adopted published values for each target. If multiple values were found, we adopted the measurement made using the highest-resolution spectrograph. Only 1 object ($\beta$ Leonis) had a $v\sin i$ reported without uncertainties; we assume $10\%$ error bars on this measurement, typical for the uncertainties in our sample. 2 objects (GJ 581 and HD 23484) had upper limits on their projected rotational velocities and were treated as such in our analysis. We note that spectral line broadening from rotation is degenerate with turbulence in the stellar atmosphere and the $v\sin i$ measurements in this work use a variety of modeling frameworks to account for macroturbulence, possibly introducing unknown systematics to our analysis.

\begin{figure}
    \centering
    \includegraphics[width=0.5\linewidth]{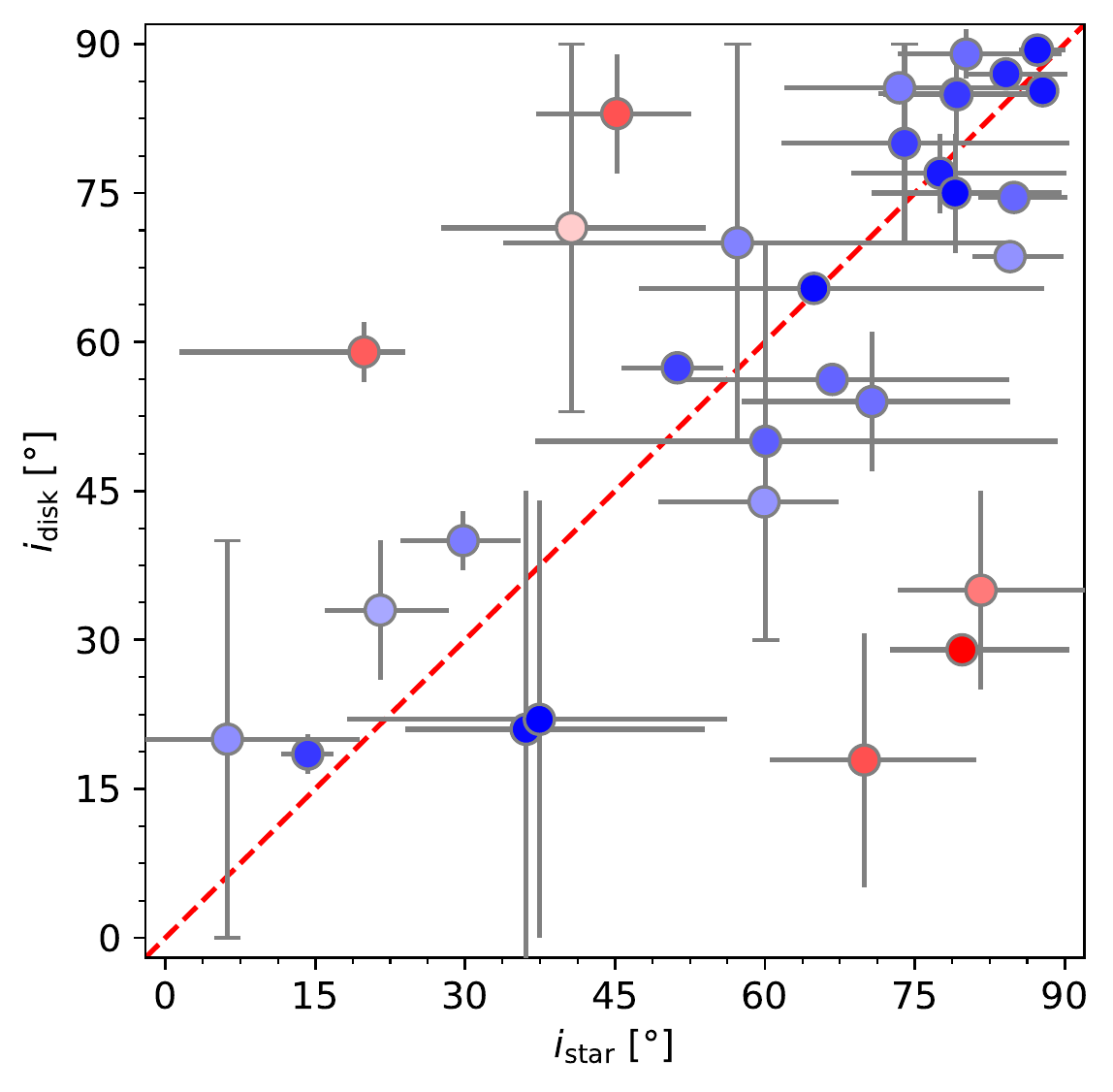}
    \caption{Disk inclinations plotted against the stellar inclinations for each system in our sample. The color of each point corresponds to the absolute value of the median of $\Delta i = i_d-i_s$, where blue indicates a well-aligned system and red indicates a large misalignment. Capped error bars represent the range of values that a parameter falls within (assuming a uniform distribution) while the rest represent the $68\%$ credibility interval.}
    \label{fig:diskvstar}
\end{figure}

Archival rotation periods were gathered for 26 objects in our sample. We also directly measured the rotation period for stars that displayed quasiperiodic variations in the Pre-Search Data Conditioned Simple Aperture Photometry (PDCSAP) light curves produced by the \textit{TESS} Science Processing Operations Center (SPOC), which have been corrected for instrumental systematics \citep{stumpe:2012, Smith:2012, Stumpe:2014}. Each photometric time series was modeled using a Gaussian process (GP), which are commonly used to represent rotational modulation induced by active regions rotating in and out of view \citep{Haywood:2014, Rajpaul:2015}. We used the rotation kernel implemented in {\tt celerite2} that combines two dampened simple harmonic oscillators with periods of $P$ and $P/2$ to capture the stochastic variability in a star's rotation signal \citep{celerite1, celerite2}. Using \textit{TESS} data, we measured the rotation period of 18 stars, 5 of which had no previously published measurements; all of these measurements agree with either our archival values or the rotation period relationship in \cite{1984ApJ...279..763N}. For each of these 18 targets, we used the rotation periods and uncertainties determined using \textit{TESS} data as they are well-constrained and measured under a standard framework. These periods, along with the projected rotational velocities and the corresponding uncertainties, can be found in \rftabl{sample} while the \textit{TESS} light curves, GP models, and rotation period posteriors are shown in Appendix \ref{sec:lightcurves}.

We then determined the stellar inclination for each target with a known radius, rotation period, and $v\sin i$ using the projected rotational velocity method, where the inclination is given by
\begin{equation}
    i = \arcsin\left(
    \frac{v\sin i}{v}\right) = \arcsin\left(\frac{v\sin i}{2\pi R / P}\right).
\end{equation}
As discussed by \cite{Masuda:2020}, $v\sin i$ and $v$ are not independent from each other, complicating the statistical inference of $i$. A simple technique accounting for this is to to use a Markov chain Monte Carlo (MCMC) process with a uniform prior on $\cos i$ and measurement informed priors on $R$, $P$, and $\left(2\pi R/P\right)\sqrt{1-\cos^2 i}$ \citep{Albrecht:2022}. This approach is also advantageous because it easily accounts for uncertainties in our measurements of $R$, $P$, and $v\sin i$. Given that measurements of $v\sin i$ are typically made from a star's spectral absorption lines and require that broadening from rotation be distinguished from other sources, including turbulence in the stellar atmosphere or instrumental resolution, the projected rotational velocity method is often subject to systematic uncertainties. Therefore, we adopted stellar inclinations previously determined using more accurate methods such as interferometry (Vega and $\beta$ Leonis), asteroseismology ($\beta$ Pictoris), and starspot tracking ($\epsilon$ Eridani) whenever possible. Interferometry and asteroseismology also allow us to expand our sample to early-type stars with weak, often undetectable rotational modulation.

We conducted a literature search for disk inclinations ($i_d$), selecting values that were the most well-constrained, typically corresponding to images with the highest spatial resolution. Most of these images were taken using ALMA, HST, and GPI, although the uncertainties on the inclinations vary widely as the spatial resolution is highly dependent on instrument configuration and distance. Inclinations can also be determined more precisely for edge-on disks than face-on disks.

To better understand whether the star and disk might be misaligned, we calculated the difference between the disk and stellar inclinations ($\Delta i = i_d - i_s$), the absolute value of which gives the minimum star-disk misalignment; because we are unable to determine the position angle of the stellar rotation axis or the direction of the disk and stellar angular momenta, we are unable to calculate the full obliquity. For systems with stellar inclinations determined using the projected rotational velocity technique, we assumed the MCMC posterior distribution; for the systems with archival measurements, a sample of stellar inclinations were drawn from Gaussian distributions. Similarly, we drew a sample of disk inclinations using either uniform or Gaussian distributions when appropriate. We then took the differences between our samples of disk and stellar inclinations and adopted the median value along with lower and upper uncertainties representative of the $68\%$ credibility interval. These differences, along with the stellar and disk inclinations, are given in \rftabl{inc}.

\begin{figure}
    \centering
    \includegraphics[width=\linewidth]{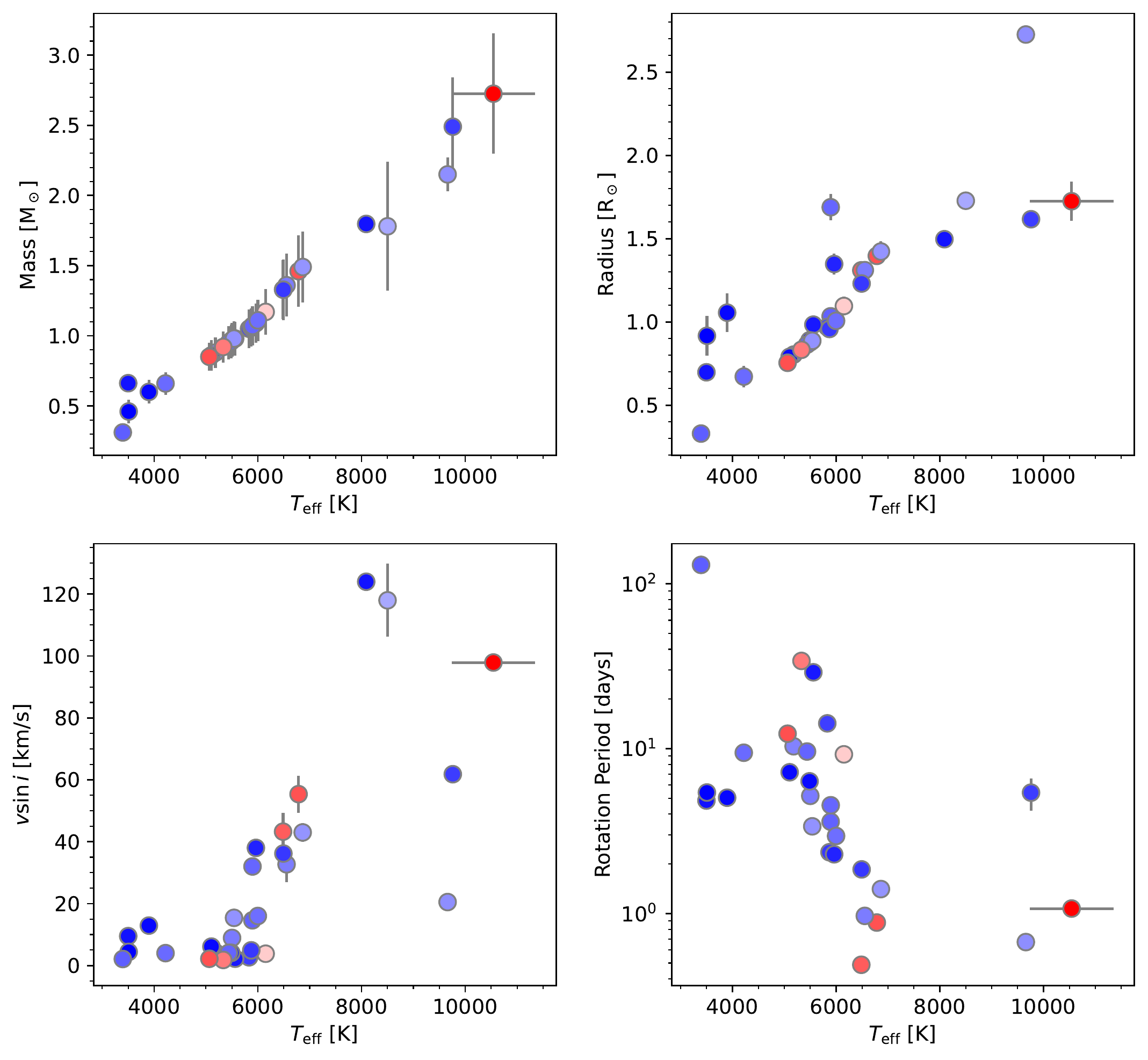}
    \caption{Stellar parameters for the systems in our sample. The color of each point corresponds to the absolute value of the median of $\Delta i = i_d-i_s$, where blue indicates a well-aligned system and red indicates a large misalignment. The top left figure shows mass versus $T_\mathrm{eff}$, the top right shows radius versus $T_\mathrm{eff}$, the bottom left gives $v\sin i$ versus $T_\mathrm{eff}$, and the bottom right gives the rotation period versus $T_\mathrm{eff}$.}
    \label{fig:params}
\end{figure}

\section{Results and Discussion}
\label{sec:results}

\subsection{Comparing Disk and Stellar Inclinations}
\label{sec:inclinations}

25 systems appear to be closely aligned with disk and stellar inclinations consistent of being within $10\degree$ of each other (although large uncertainties mean that some of these systems could still be misaligned). There are several exceptions; most notably, HD 10647, HD 138813, HD 191089, HD 30447, $\epsilon$ Eridani, and $\tau$ Ceti all have misalignments ranging roughly between $30\degree$ and $60\degree$. If stars and their disks are well-aligned, we would expect to see a monotonic, increasing relationship between disk inclination and stellar inclination in \rffigl{diskvstar}. We test how well-aligned systems tend to be by calculating the Spearman rank-order correlation coefficient ($r_S$) for our data set. Using the median values for our inclinations, we find a coefficient of 0.62 with a p-value of 0.0002; however, this does not reflect the broad uncertainties on some of the inclination measurements. For each disk and stellar inclination, we drew a random sample and calculated a new coefficient and the corresponding p-value $10^4$ times. The 68\% credibility interval for $r_S$ was $.54\pm0.08$ with p-values of $0.0008_{-0.00069}^{+0.0036}$. These values for $r_S$ are notably lower than the coefficient of 0.82 found by \cite{Watson:2011} and indicate that while there is a positive correlation between disk and stellar inclinations, they are not always well-aligned.

It is important to keep in mind that disk and stellar inclinations can only put a lower limit on misalignment and that a full analysis requires knowledge of both the disk and stellar position angle on the sky plane. Further, inclinations do not indicate the directions that the star is rotating and the disk material is orbiting; if they are moving in opposite directions, the misalignment between the disk and star would be much greater than calculated. Given that systems such as K2-290---a strong candidate for primordial misalignment---have co-planar planets in retrograde orbits, this may be a significant bias \citep{Hjorth:2021}.

While \cite{Watson:2011} and \cite{Greaves:2014} did not find signs of star-disk misalignment in their sample of debris disks, \cite{Davies:2019} observed misalignment of protoplanetary disks at a rate slightly higher than seen in our analysis ($\mysim33\%$); however, we note that they observed much smaller misalignments, typically less than $30\degree$. This indicates that the star-disk misalignment may not decrease as the disk transitions, as suggested by \cite{Davies:2019}, and raises the question of whether misalignment increases as the system evolves. It is possible that mechanisms such as stellar flybys can incline debris disks \citep{2020ApJ...901...92M} while processes such as accretion onto the star are unlikely to realign the system.

\begin{figure}
    \centering
    \includegraphics[width=\linewidth]{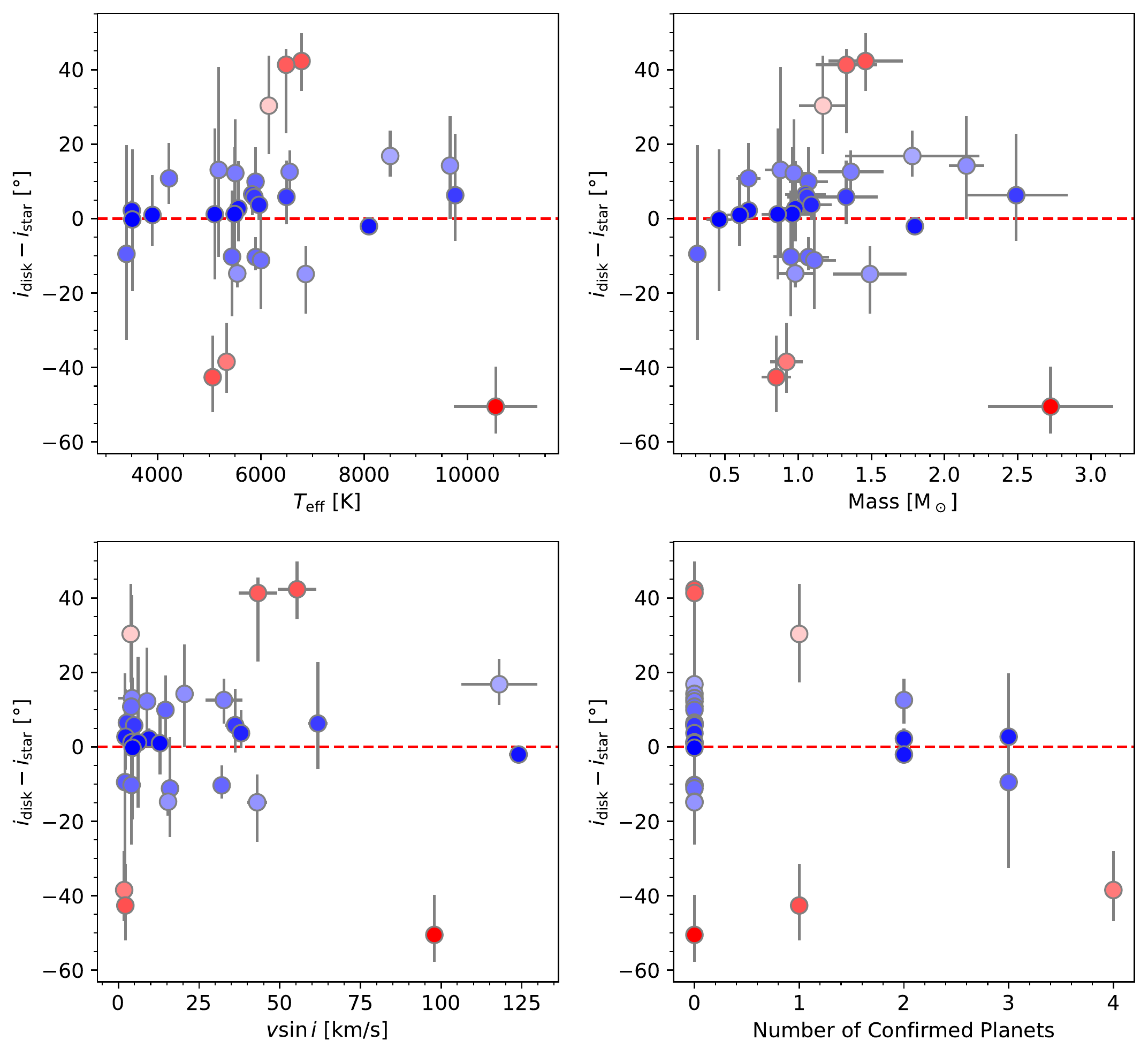}
    \caption{The difference between the disk and stellar inclinations ($i_d-i_s$) plotted versus different system parameters. The color of each point corresponds to the median value of the minimum obliquity. The top left shows $T_\mathrm{eff}$ on the bottom axis while the right has the mass, the bottom left shows $v\sin i$, and the bottom right the number of confirmed planets in the system.}
    \label{fig:diff}
\end{figure}

\rffigl{params} shows the mass, radius, $v\sin i$, and rotation period of each star in our sample versus the effective temperature. \rffigl{diff} shows the difference between the disk and stellar inclinations as a function of system parameters. In these plots, we see most of the star-disk systems are well-aligned aside from the 6 mentioned above. The misaligned systems are not clustered around any specific $T_\mathrm{eff}$ or mass, suggesting that misalignment may occur regardless of stellar type, although there are not enough stars to make definitive conclusions. We also do not observe misalignment occurring more frequently with the presence of known planets; yet, many substellar objects in these debris disk systems may easily be undetectable. Finally, the 6 significantly misaligned systems span a wide range of ages ($\mysim$7 Myr to 5.8 Gyr; \citealt{2008ApJ...687.1264M, Bell:2015, Pecaut:2016, 2017AJ....154...69S, Nielsen:2019}); this is not surprising given that primordial misalignment is expected to occur during the protoplanetary disk stage, well before debris disks form.

\subsection{Implications for Primordial Misalignment}

If magnetic warping were responsible for spin-orbit misalignment, \cite{Spalding:2015} argue that misalignment should occur more frequently around stars with masses greater than 1.2 $M_\odot$; this is because lower-mass, young stars would be able to realign their stellar spin axes with the surrounding disks due to their stronger magnetic fields. As seen in \rffigl{diff}, 2 of the significantly misaligned systems ($\epsilon$ Eridani and $\tau$ Ceti) have stellar masses below 1.2 $M_\odot$ while 2 (HD 10647 and HD 191089) have masses very close to this limit, suggesting that magnetic warping alone is not a viable mechanism for disk misalignment. 

If chaotic accretion were at play, subsequent accretion of disk material onto the star is expected to reduce the misalignment to values lower than $20\degree$ by the time planets begin to form \citep{Takaishi:2020}. Not only does this fail to describe the distribution of obliquities observed for exoplanets, but it does not match the $\mysim30\degree-60\degree$ misalignments shown in \rffigl{diff}. We do see systems with small potential misalignments near or below $20\degree$, including HD 107146, HD 129590, HD 145560, HD 202917, HD 206893, HD 35650, HD 377, and $\beta$ Leonis, but the large uncertainties on our obliquity measurements make it difficult to determine whether low-obliquity systems are truly misaligned. Additionally, without knowing the position angles of each star, we cannot definitively comment on whether star-disk misalignment commonly falls near $20\degree$. 

While the significantly misaligned disks could have been torqued out of alignment by an inclined stellar or planetary companion, this mechanism is unable to explain the observed distribution of spin-orbit obliquities \citep{Zanazzi:2018, Albrecht:2022}. Ultimately, it is unclear what mechanisms can misalign disks around their stars; further, because we do not know of many planets in these systems, we are unable to determine whether the same mechanisms could be responsible for spin-orbit misalignment. As discussed in \rfsecl{inclinations}, mechanisms such as stellar flybys may incline debris disks in addition to planetary orbits, and the obliquities measured in this work may not reflect a system's primordial architecture.

\section{Conclusions}
\label{sec:conclusions}

We investigate the alignment of resolved debris disks with their stars, placing a lower limit on their obliquities by comparing stellar and disk inclinations. With recent resolved images of disks taken by ALMA, HST, and GPI, along with rotation periods measured using \textit{TESS}, we were able to include 31 systems in our analysis, more than 3 times as large as the samples included in previous studies of debris disks.

While there formerly was little evidence for misalignment between debris disks and their stars, we find 6 systems with disk and stellar inclinations separated by $\mysim30\degree-60\degree$. This indicates that these evolved disks are frequently misaligned with their stars; although, systems are more often well-aligned than not. Given that we observe such large minimum obliquities, some mechanism other than chaotic accretion needs to be at play. We also see misaligned systems with stellar masses below or near 1.2 $M_\odot$, suggesting that magnetic warping alone cannot be responsible for misalignment. Because resonant processes that could torque the disk out of alignment fail to explain the distribution of spin-orbit obliquities, it remains unclear what role primordial misalignment could play in shaping planetary systems. Further, it is unknown whether these disk obliquities truly reflect the structure of the preceding protoplanetary disk.

Future work needs to expand the number of debris disk hosts with inclination measurements, helping constrain the characteristics of misaligned systems. Few stars in our sample have known planetary companions and no confirmed hot Jupiter systems are currently known to contain circumstellar debris; searching for dust in confirmed planetary systems could help better understand whether the mechanisms that misalign disks with their stars are also responsible for spin-orbit misalignment. 

Existing methods to measure stellar position angle cannot be applied to the vast majority of debris disk hosts \citep{LeBouquin:2009, Lesage:2014}, meaning the full obliquity between a disk and its star cannot be measured. As mentioned by \cite{Watson:2011}, a full Bayesian analysis accounting for this limitation could place more useful upper limits on the misalignment, similar to the framework presented in \cite{Fabrycky:2009} for spin-orbit angles. Regardless, the lower limits on misalignment presented in this work help better understand the geometry of debris disks, and future observations will improve our understanding of the mechanisms that shape and misalign these systems.


\begin{acknowledgments}
We thank the referee for thorough and insightful feedback that greatly improved the quality of this paper. We also thank Ruth Angus and Megan Bedell for a helpful discussion on how to measure stellar rotation periods and Ann-Marie Madigan and Carolyn Crow for thoughtful conversation about our results.
M.A.M. acknowledges support for this work from the National Aeronautics and Space Administration (NASA) under award number 19-ICAR19\_2-0041.  This research has made use of the NASA Exoplanet Archive, which is operated by the California Institute of Technology, under contract with the National Aeronautics and Space Administration under the Exoplanet Exploration Program.
This work made use of the SIMBAD database (operated at CDS, Strasbourg, France), NASA’s Astrophysics Data System Bibliographic Services.
The TIC data presented in this paper were obtained from the Mikulski Archive for Space Telescopes (MAST) at the Space Telescope Science Institute and can be accessed via
\dataset[10.17909/fwdt-2x66]{https://doi.org/10.17909/fwdt-2x66}.
This research has made use of the VizieR catalog access tool, CDS, Strasbourg, France (DOI: 10.26093/cds/vizier). The original description of the VizieR service was published in A$\&$AS 143, 23.
This work has made use of data from the European Space Agency (ESA) mission {\it Gaia} (\url{https://www.cosmos.esa.int/gaia}), processed by the {\it Gaia} Data Processing and Analysis Consortium (DPAC, \url{https://www.cosmos.esa.int/web/gaia/dpac/consortium}). Funding for the DPAC has been provided by national institutions, in particular the institutions participating in the {\it Gaia} Multilateral Agreement.
\end{acknowledgments}

\software{{\tt astropy} \citep{astropy:2018}, {\tt exoplanet} \citep{exoplanet:2021}, {\tt matplotlib} \citep{Hunter:2007}, {\tt numpy} \citep{Harris:2020}, {\tt PyMC3} \citep{pymc3}, {\tt SciPy} \citep{scipy}}

\clearpage
\begin{longrotatetable}
\begin{deluxetable*}{@{\extracolsep{0.5pt}}lccccccccccccccc}
\tablecaption{Sample of Disk Hosts and Stellar Properties \label{tab:sample}}
\tabletypesize{\small}
\tablehead
{
\colhead{}& \colhead{}& \colhead{}& \colhead{}& \colhead{}& \colhead{}& \colhead{}& \colhead{}& \multicolumn{6}{c}{References} \\
\cline{9-15}
\colhead{Name}& \colhead{SpT}& \colhead{$T_\mathrm{eff}$}& \colhead{Mass}& \colhead{Radius}& \colhead{$v\sin i$}& \colhead{$P_\mathrm{rot}$}& \colhead{$N_\mathrm{planets}$}& \colhead{SpT}& \colhead{$T_\mathrm{eff}$}& \colhead{Mass}& \colhead{Radius}& \colhead{$v\sin i$}& \colhead{$P_\mathrm{rot}$} \\[-1.5ex]
\colhead{}& \colhead{}& \colhead{(K)}& \colhead{($M_\odot$)}& \colhead{($R_\odot$)}& \colhead{($\mathrm{km\,s^{-1}}$)}& \colhead{(days)}& \colhead{}& \colhead{}& \colhead{}& \colhead{}& \colhead{}
}
\startdata
61 Virginis & G6.5V & $5562\pm110$ & $0.98\pm0.12$ & $0.984\pm0.045$ & $2.2\pm0.3$ & 29.0 & 3\tablenotemark{a} & 1 & 2 & 2 & 2 & 3 & 4 \\ 
AU Mic & M1VeBa1 & $3500\pm100$ & $0.66\pm0.02$ & $0.698\pm0.021$ & $9.5\pm0.2$ & $4.8366\pm0.0075$ & 2\tablenotemark{b} & 1 & 5 & 2 & 2 & 6 & 7 \\ 
GJ 581 & M3V & $3396\pm160$ & $0.31\pm0.02$ & $0.33\pm0.01$ & $\leq2.1$ & $130\pm2$ & 3\tablenotemark{c} & 8 & 2 & 2 & 2 & 9 & 10 \\ 
HD 104860 & G0/F9V & $6000\pm150$ & $1.11\pm0.15$ & $1.007\pm0.054$ & $16\pm2$ & $2.9493\pm0.0031$ & 0 & 11 & 2 & 2 & 2 & 11 & 7 \\ 
HD 10647 & F9V & $6151\pm150$ & $1.17\pm0.16$ & $1.096\pm0.057$ & $3.8\pm0.4$ & $9.22\pm0.06$ & 1\tablenotemark{d} & 1 & 2 & 2 & 2 & 12 & 7 \\ 
HD 107146 & G2V & $5873\pm110$ & $1.06\pm0.14$ & $0.958\pm0.041$ & $4.9\pm0.6$ & 2.35 & 0 & 13 & 2 & 2 & 2 & 14 & 15 \\ 
HD 129590 & G3V & $5897\pm130$ & $1.07\pm0.14$ & $1.689\pm0.079$ & $32\pm2$ & $4.5291\pm0.0061$ & 0 & 16 & 2 & 2 & 2 & 17 & 7 \\ 
HD 138813 & A0V & $10543\pm810$ & $2.72\pm0.43$ & $1.72\pm0.12$ & $97.9\pm2.2$ & 1.07 & 0 & 18 & 2 & 2 & 2 & 19 & 20 \\ 
HD 141943 & G2 & $5963\pm120$ & $1.09\pm0.14$ & $1.348\pm0.062$ & $38\pm2$ & $2.283\pm0.062$ & 0 & 21 & 2 & 2 & 2 & 22 & 7 \\ 
HD 145560 & F5V & $6865\pm140$ & $1.49\pm0.25$ & $1.423\pm0.061$ & $43\pm3$ & $1.40473\pm0.00061$ & 0 & 23 & 2 & 2 & 2 & 17 & 7 \\ 
HD 166 & G8 & $5489\pm140$ & $0.96\pm0.12$ & $0.89\pm0.052$ & $4.1\pm0.4$ & $6.3265\pm0.0082$ & 0 & 24 & 2 & 2 & 2 & 25 & 7 \\ 
HD 181296 & A0V & $9760\pm140$ & $2.49\pm0.35$ & $1.617\pm0.042$ & $62\pm3$ & $5.4\pm1.2$ & 0 & 26 & 2 & 2 & 2 & 27 & 27 \\ 
HD 191089 & F5V & $6487\pm110$ & $1.33\pm0.21$ & $1.31\pm0.053$ & $43\pm6$ & $0.488\pm0.005$ & 0 & 16 & 2 & 2 & 2 & 27 & 28 \\ 
HD 202628 & G5V & $5831\pm130$ & $1.05\pm0.14$ & $0.97\pm0.05$ & $2.65\pm0.11$ & $14.215\pm0.091$ & 0 & 21 & 2 & 2 & 2 & 29 & 7 \\ 
HD 202917 & G7V & $5541\pm140$ & $0.98\pm0.12$ & $0.887\pm0.052$ & $15.4\pm0.1$ & $3.3739\pm0.0036$ & 0 & 21 & 2 & 2 & 2 & 21 & 7 \\ 
HD 206893 & F5V & $6554\pm120$ & $1.36\pm0.22$ & $1.311\pm0.055$ & $32.7\pm5.7$ & $0.966\pm0.003$ & 2\tablenotemark{e} & 30 & 2 & 2 & 2 & 31 & 32 \\ 
HD 23484 & K2V & $5182\pm140$ & $0.88\pm0.11$ & $0.804\pm0.054$ & $\leq4.2$ & $10.32\pm0.83$ & 0 & 21 & 2 & 2 & 2 & 21 & 7 \\ 
HD 30447 & F3V & $6786\pm110$ & $1.46\pm0.25$ & $1.397\pm0.053$ & $55\pm6$ & $0.88\pm0.02$ & 0 & 16 & 2 & 2 & 2 & 27 & 27 \\ 
HD 35650 & K6V & $4220\pm130$ & $0.66\pm0.081$ & $0.672\pm0.063$ & $4.0\pm0.2$ & $9.43\pm0.03$ & 0 & 21 & 2 & 2 & 2 & 21 & 7 \\ 
HD 35841 & F3V & $6493\pm130$ & $1.33\pm0.22$ & $1.231\pm0.051$ & $36\pm3$ & $1.849\pm0.019$ & 0 & 18 & 2 & 2 & 2 & 27 & 7 \\ 
HD 377 & G2V & $5895\pm140$ & $1.07\pm0.13$ & $1.034\pm0.055$ & $14.6\pm0.5$ & $3.601\pm0.046$ & 0 & 21 & 2 & 2 & 2 & 25 & 7 \\ 
HD 53143 & G9V & $5440\pm100$ & $0.95\pm0.12$ & $0.864\pm0.039$ & $4\pm1$ & $9.592\pm0.027$ & 0 & 21 & 2 & 2 & 2 & 33 & 7 \\ 
HD 61005 & G8Vk & $5503\pm110$ & $0.97\pm0.12$ & $0.878\pm0.039$ & $8.9\pm1.5$ & $5.1639\pm0.0078$ & 0 & 30 & 2 & 2 & 2 & 27 & 7 \\ 
HD 92945 & K1V & $5105\pm150$ & $0.86\pm0.11$ & $0.791\pm0.056$ & $6\pm3$ & $7.18\pm0.31$ & 0 & 21 & 2 & 2 & 2 & 34 & 7 \\ 
TWA 25 & M0.5 & $3899\pm120$ & $0.6\pm0.084$ & $1.06\pm0.12$ & $12.9\pm1.2$ & $5.0318\pm0.0079$ & 0 & 35 & 2 & 2 & 2 & 36 & 7 \\ 
TWA 7 & M2Ve & $3509\pm120$ & $0.46\pm0.086$ & $0.92\pm0.12$ & $4.4\pm1.2$ & $5.41\pm0.37$ & 0 & 21 & 2 & 2 & 2 & 21 & 7 \\ 
Vega & A0Va & $9660\pm90$ & $2.15\pm0.12$ & $2.726\pm0.006$ & $20.5\pm0.7$ & $0.67\pm0.07$ & 0 & 37 & 38 & 38 & 38 & 39 & 40 \\ 
$\beta$ Leonis & A3Va & $8502\pm45$ & $1.93\pm0.01$ & $2.28\pm0.11$ & $118\pm12$ & \nodata & 0 & 41 & 2 & 42 & 42 & 43 & \nodata \\ 
$\beta$ Pictoris & A6V & $8143\pm67$ & $1.8\pm0.04$ & $1.54\pm0.04$ & $124\pm3$ & \nodata & 2\tablenotemark{f} & 30 & 44 & 44 & 44 & 45 & \nodata \\ 
$\epsilon$ Eridani & K2V & $5065\pm100$ & $0.9\pm0.1$ & $0.755\pm0.038$ & $2.17\pm0.25$ & $12.3\pm0.06$ & 1\tablenotemark{g} & 1 & 2 & 2 & 2 & 6 & 46 \\ 
$\tau$ Ceti & G8V & $5333\pm99$ & $0.92\pm0.11$ & $0.833\pm0.037$ & $1.8\pm0.1$ & 34.0 & 4\tablenotemark{h} & 1 & 2 & 2 & 2 & 3 & 47 \\
\enddata

\tablenotetext{a}{Planets around 61 Virginis detected via radial velocities and inclinations are unknown \citep{2010ApJ...708.1366V}.}
\tablenotetext{b}{Planets follow edge-on orbits around AU Mic and are well-aligned with star and disk \citep{2020Natur.582..497P, 2021AJ....162..137A, 2021AA...649A.177M}.}
\tablenotetext{c}{Planets around GJ 581 detected via radial velocities and inclinations are unknown \citep{2018AA...609A.117T}.}
\tablenotetext{d}{Planet around HD 10647 detected via radial velocities and inclination is unknown \citep{2013AA...551A..90M}.}
\tablenotetext{e}{The two planets around HD 206893 have a mutual inclination of 15$\degree$ at most and could be misaligned from the disk by $\mysim20\degree$ \citep{2022arXiv220804867H}.
}
\tablenotetext{f}{Planets around $\beta$ Pictoris detected via direct imaging and appear well-aligned with the disk \citep{2022ApJS..262...21F}.}
\tablenotetext{g}{Orbital inclination of planet around $\epsilon$ Eridani has large uncertainties, may be misaligned with the disk \citep{2021AJ....162..181L, 2022RNAAS...6...45B}.}
\tablenotetext{h}{Planets around $\tau$ Ceti detected via radial velocities and inclinations are unknown \citep{2017AJ....154..135F}.}

\tablerefs{(1) \cite{1989ApJS...71..245K}, (2) 2021arXiv210804778P, (3) \cite{2003AA...398..647R}, (4) \cite{1996ApJ...457L..99B}, (5) \cite{2009ApJ...698.1068P}, (6) \cite{2019AA...629A..80H}, (7) This Work, (8) \cite{1986AJ.....92..139S}, (9) \cite{2005ESASP.560..571G}, (10) \cite{2014Sci...345..440R}, (11) \cite{2016MNRAS.458.2307K}, (12) \cite{2018AA...615A..76S}, (13) \cite{1970AJ.....75..507H}, (14) \cite{2016ApJS..225...32B}, (15) \cite{2012MNRAS.427.2917R}, (16) \cite{1982mcts.book.....H}, (17) \cite{2011ApJ...738..122C}, (18) \cite{1988mcts.book.....H}, (19) \cite{2012ApJ...745...56D}, (20) \cite{2018AJ....155...39O}, (21) \cite{2006AA...460..695T}, (22) \cite{2002AA...384..491C}, (23) \cite{1978mcts.book.....H}, (24) \cite{1985ApJS...59..197B}, (25) \cite{2014MNRAS.444.3517M}, (26) \cite{1975mcts.book.....H}, (27) \cite{2021AA...645A..30Z}, (28) \cite{2015AA...573A.126D}, (29) \cite{2016AA...592A.156D}, (30) \cite{2006AJ....132..161G}, (31) \cite{2022AA...667A..63Z}, (32) \cite{2017AA...608A..79D}, (33) \cite{1997MNRAS.284..803S}, (34) \cite{2000AAS..142..275S}, (35) \cite{2014ApJ...786...97H}, (36) \cite{2014ApJ...788...81M}, (37) \cite{2003AJ....126.2048G}, (38) \cite{2012ApJ...761L...3M}, (39) \cite{2021AA...647A..49S}, (40) \cite{2021AJ....161..157H}, (41) \cite{2009ApJ...694.1085V}, (42) \cite{2004AA...426..601D}, (43) \cite{2009ApJ...691.1896A}, (44) \cite{2019AA...627A..28Z}, (45) \cite{2003MNRAS.344.1250K}, (46) \cite{2016ApJ...824..150G}, (47) \cite{2010ApJ...725..875I}}

\end{deluxetable*}
\end{longrotatetable}

\startlongtable
\begin{deluxetable*}{@{\extracolsep{3pt}}lccccccc}
\tablecaption{Disk and Stellar Inclinations\label{tab:inc}}
\tabletypesize{\footnotesize}
\tablehead
{
\colhead{}& \colhead{}& \colhead{}& \colhead{}& \colhead{}& \colhead{}& \multicolumn{2}{c}{References} \\
\cline{7-8}
\colhead{Name}& \colhead{$i_\mathrm{disk}$}& \colhead{Disk Imaging Facility}& \colhead{$i_\mathrm{star}$}& \colhead{Stellar Inclination Method}& \colhead{$\Delta i$}& \colhead{$i_\mathrm{disk}$}& \colhead{$i_\mathrm{star}$} \\
\colhead{}& \colhead{($\degree$)}& \colhead{}& \colhead{($\degree$)}& \colhead{}& \colhead{($\degree$)}& \colhead{}& \colhead{}
}
\startdata
61 Virginis & $77\pm4$ & HSO & $77.5^{+8.7}_{-13.0}$ & $v\sin i$ & $2.7^{+13.0}_{-8.9}$ & 1 & 2 \\ 
AU Mic & $89.4\pm0.1$ & GPI & $87.3^{+1.9}_{-2.8}$ & $v\sin i$ & $2.2^{+2.8}_{-1.9}$ & 3 & 2 \\ 
GJ 581 & [30.0,70.0] & HSO & $60.0^{+21.0}_{-28.0}$ & $v\sin i$ & $-9.0^{+29.0}_{-23.0}$ & 4 & 2 \\ 
HD 104860 & $54\pm7$ & HSO & $71.0\pm13.0$ & $v\sin i$ & $-11.0^{+14.0}_{-13.0}$ & 5 & 2 \\ 
HD 10647 & [53.0,90.0] & HSO & $40.6^{+6.4}_{-5.6}$ & $v\sin i$ & $30.0^{+14.0}_{-13.0}$ & 6 & 2 \\ 
HD 107146 & $18\pm2$ & HST & $14.2^{+2.4}_{-2.3}$ & $v\sin i$ & $5.8^{+2.6}_{-2.7}$ & 7 & 2 \\ 
HD 129590 & $74.56\pm0.05$ & SPHERE/VLT & $84.9^{+3.6}_{-5.4}$ & $v\sin i$ & $-10.3^{+5.4}_{-3.6}$ & 8 & 2 \\ 
HD 138813 & $29.0\pm0.3$ & ALMA & $79.7^{+7.3}_{-11.0}$ & $v\sin i$ & $-50.5^{+11.0}_{-7.3}$ & 9 & 2 \\ 
HD 141943 & $87\pm1$ & GPI & $84.1^{+4.1}_{-6.2}$ & $v\sin i$ & $3.7^{+6.2}_{-4.1}$ & 3 & 2 \\ 
HD 145560 & $43.9\pm1.5$ & GPI & $59.9^{+11.0}_{-7.5}$ & $v\sin i$ & $-14.9^{+7.5}_{-11.0}$ & 10 & 2 \\ 
HD 166 & $21\pm24$ & HSO & $36.1^{+5.4}_{-4.5}$ & $v\sin i$ & $1.0^{+18.0}_{-12.0}$ & 5 & 2 \\ 
HD 181296 & [70.0,90.0] & Gemini South & $74.0^{+11.0}_{-16.0}$ & $v\sin i$ & $6.0^{+17.0}_{-12.0}$ & 11 & 2 \\ 
HD 191089 & $59\pm3$ & HST & $19.9^{+20.0}_{-3.6}$ & $v\sin i$ & $41.3^{+4.2}_{-19.0}$ & 12 & 2 \\ 
HD 202628 & $57.4\pm0.4$ & ALMA & $51.2^{+5.5}_{-4.6}$ & $v\sin i$ & $6.5^{+4.6}_{-5.6}$ & 13 & 2 \\ 
HD 202917 & $68.6\pm1.5$ & HST & $84.5^{+3.8}_{-5.4}$ & $v\sin i$ & $-14.7^{+5.4}_{-3.8}$ & 14 & 2 \\ 
HD 206893 & $40\pm3$ & ALMA & $29.8^{+6.0}_{-5.6}$ & $v\sin i$ & $12.6^{+5.8}_{-6.3}$ & 15 & 2 \\ 
HD 23484 & [50.0,90.0] & Herschel & $57.0^{+22.0}_{-26.0}$ & $v\sin i$ & $13.0^{+28.0}_{-23.0}$ & 16 & 2 \\ 
HD 30447 & $83\pm6$ & GPI & $45.2^{+7.6}_{-6.2}$ & $v\sin i$ & $42.3^{+7.5}_{-8.1}$ & 3 & 2 \\ 
HD 35650 & $89.0\pm2.5$ & HST & $80.1^{+6.9}_{-9.4}$ & $v\sin i$ & $10.8^{+9.6}_{-6.9}$ & 17 & 2 \\ 
HD 35841 & $84.9\pm0.2$ & GPI & $79.2^{+7.4}_{-9.7}$ & $v\sin i$ & $5.8^{+9.7}_{-7.4}$ & 18 & 2 \\ 
HD 377 & $85$\tablenotemark{a} & HST & $79.1^{+7.4}_{-8.9}$ & $v\sin i$ & $9.9^{+9.3}_{-7.8}$ & 17 & 2 \\ 
HD 53143 & $56.23\pm0.37$ & ALMA & $67.0^{+16.0}_{-18.0}$ & $v\sin i$ & $-10.0^{+18.0}_{-16.0}$ & 19 & 2 \\ 
HD 61005 & $85.6\pm0.1$ & ALMA & $73.0^{+12.0}_{-14.0}$ & $v\sin i$ & $12.0^{+14.0}_{-12.0}$ & 20 & 2 \\ 
HD 92945 & $65.4\pm0.9$ & ALMA & $65.0^{+18.0}_{-23.0}$ & $v\sin i$ & $1.0^{+23.0}_{-18.0}$ & 21 & 2 \\ 
TWA 25 & $75\pm6$ & HST & $79.0^{+7.6}_{-10.0}$ & $v\sin i$ & $1.0^{+11.0}_{-8.4}$ & 17 & 2 \\ 
TWA 7 & $22\pm22$ & HST & $37.0^{+17.0}_{-11.0}$ & $v\sin i$ & $0.0\pm19.0$ & 17 & 2 \\ 
Vega & [0.0,40.0] & ALMA & $6.2\pm0.4$ & Interferometry & $14.0^{+13.0}_{-14.0}$ & 22 & 23 \\ 
$\beta$ Leonis & $33\pm7$ & HSO & $21.5\pm5.0$ & Interferometry & $16.8^{+6.9}_{-5.6}$ & 24 & 25 \\ 
$\beta$ Pictoris & $85.3\pm0.3$ & GPI & $87.8\pm1.6$ & Asteroseismology & $-2.0^{+1.6}_{-1.5}$ & 3 & 26 \\ 
$\epsilon$ Eridani & $18\pm13$ & SMA/ATCA & $69.95^{+5.6}_{-7.6}$ & Photometry + Spectroscopy & $-42.6^{+11.0}_{-9.5}$ & 27 & 28 \\ 
$\tau$ Ceti & $35\pm10$ & HSO & $81.6^{+5.9}_{-8.6}$ & $v\sin i$ & $-38.4^{+11.0}_{-8.3}$ & 29 & 2 \\ 
\enddata

\tablenotetext{a}{No uncertainties reported. Standard deviation of $5\degree$ assumed.}

\tablerefs{(1) \cite{2012MNRAS.424.1206W}, (2) This Work, (3) \cite{2020AJ....160...24E}, (4) \cite{2012AA...548A..86L}, (5) \cite{2016ApJ...831...97M}, (6) \cite{2010AA...518L.132L}, (7) \cite{2014AJ....148...59S}, (8) \cite{2017ApJ...843L..12M}, (9) \cite{2019ApJ...878..113H}, (10) \cite{2020AJ....159...31H}, (11) \cite{2009AA...493..299S}, (12) \cite{2019ApJ...882...64R}, (13) \cite{2019AJ....158..162F}, (14) \cite{2016AJ....152...64S}, (15) \cite{2020MNRAS.498.1319M}, (16) \cite{2014AA...561A.114E}, (17) \cite{2016ApJ...817L...2C}, (18) \cite{2018AJ....156...47E}, (19) \cite{2022ApJ...933L...1M}, (20) \cite{2018ApJ...869...75M}, (21) \cite{2019MNRAS.484.1257M}, (22) \cite{2020ApJ...898..146M}, (23) \cite{2012ApJ...761L...3M}, (24) \cite{2011MNRAS.417.1715C}, (25) \cite{2009ApJ...691.1896A}, (26) \cite{2019AA...627A..28Z}, (27) \cite{2015ApJ...809...47M}, (28) \cite{2016ApJ...824..150G}, (29) \cite{2014MNRAS.444.2665L}}

\end{deluxetable*}

\appendix
\section{\textit{TESS} Light Curves and Measured Rotation Periods}
\label{sec:lightcurves}

\begin{figure}[!t]
    \centering
    \includegraphics[width=\linewidth]{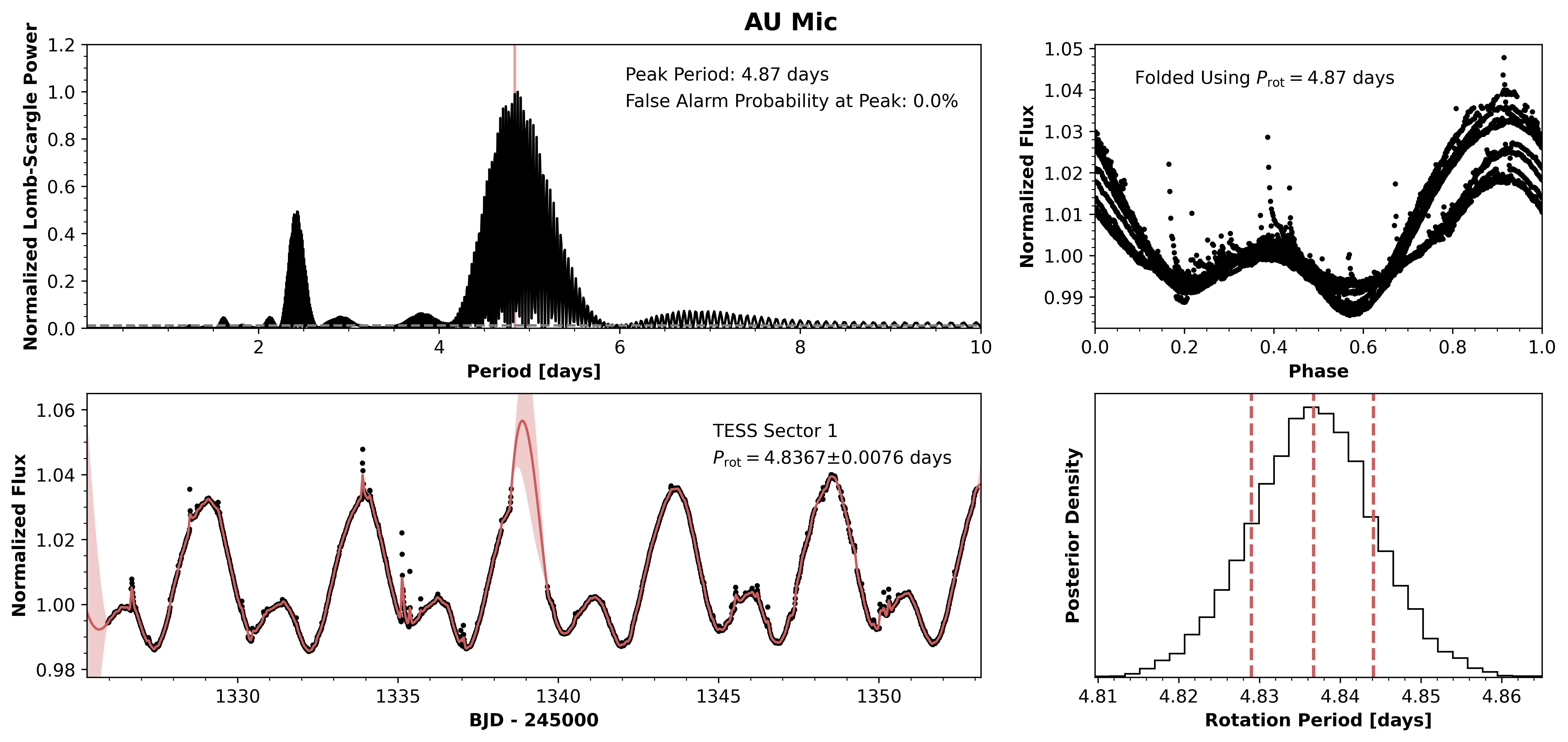}
    \caption{\textit{Top left:} The Lomb-Scargle periodogram for the AU Mic \textit{TESS} PDCSAP light curve is shown in black. The dashed grey line marks the 1$\%$ FAP level while the shaded red region denotes the $1\sigma$ confidence interval for the rotation period posterior. \textit{Top right:} The phase-folded light curve using the peak period from the Lomb-Scargle periodogram. \textit{Bottom left:} The black points show data from \textit{TESS} Sector 1 while the red line and shaded region mark the mean and $1\sigma$ confidence interval for the GP model. \textit{Bottom right:} The histogram shows the rotation period posterior derived from the GP model while the dash red lines mark the median and $1\sigma$ interval.}
    \label{fig:AUMic}
\end{figure}

\begin{figure}[!h]
    \centering
    \includegraphics[width=\linewidth]{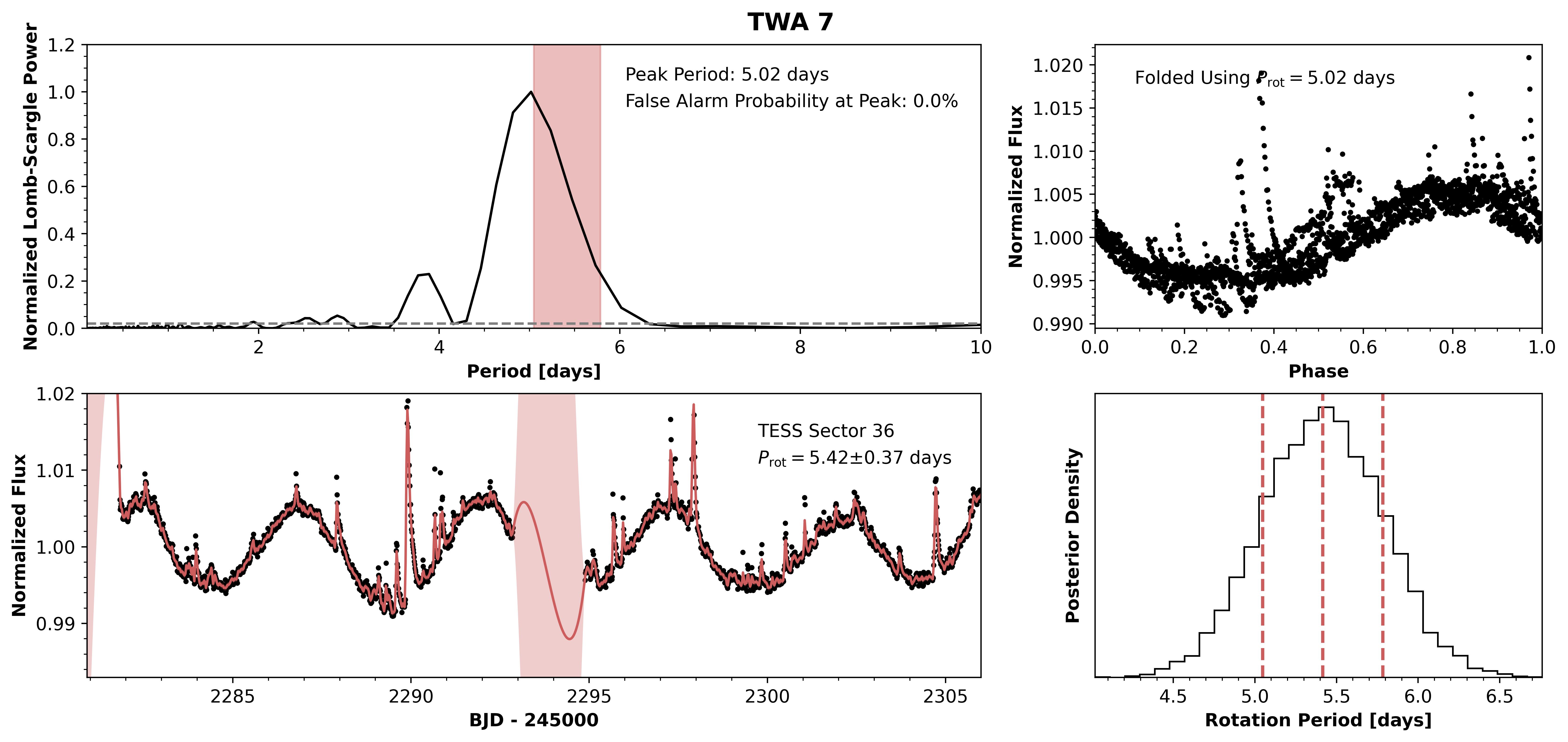}
    \caption{\textit{Top left:} The Lomb-Scargle periodogram for the TWA 7 \textit{TESS} PDCSAP light curve is shown in black. The dashed grey line marks the 1$\%$ FAP level while the shaded red region denotes the $1\sigma$ confidence interval for the rotation period posterior. \textit{Top right:} The phase-folded light curve using the peak period from the Lomb-Scargle periodogram. \textit{Bottom left:} The black points show data from \textit{TESS} Sector 36 while the red line and shaded region mark the mean and $1\sigma$ confidence interval for the GP model. \textit{Bottom right:} The histogram shows the rotation period posterior derived from the GP model while the dash red lines mark the median and $1\sigma$ interval.}
    \label{fig:TWA7}
\end{figure}

A description of the light curve modeling approach is given in \rfsecl{methods} while the derived rotation periods and $1\sigma$ uncertainties are found in \rftabl{sample}. We additionally calculate Lomb-Scargle periodograms for each light curve and the false alarm probability (FAP) associated with the rotation signal \citep{Zechmeister:2009}. Several stars display double dipping, where two opposing star spots create a false signal at $P/2$ \citep{2018ApJ...863..190B}; in several instances, including HD 377 and HD 92945, we find that phase dispersion minimization periodograms \citep{1978ApJ...224..953S} better capture the true rotation period and use them in place of Lomb-Scargle periodograms. Figures \ref{fig:AUMic} through \ref{fig:TWA7} show a periodogram, phase-folded light curve, light curve from a single \textit{TESS} sector, and the rotation period posterior for each object showing quasiperiodic variations in \textit{TESS} data.

\clearpage

\begin{figure}[!b]
    \centering
    \includegraphics[width=\linewidth]{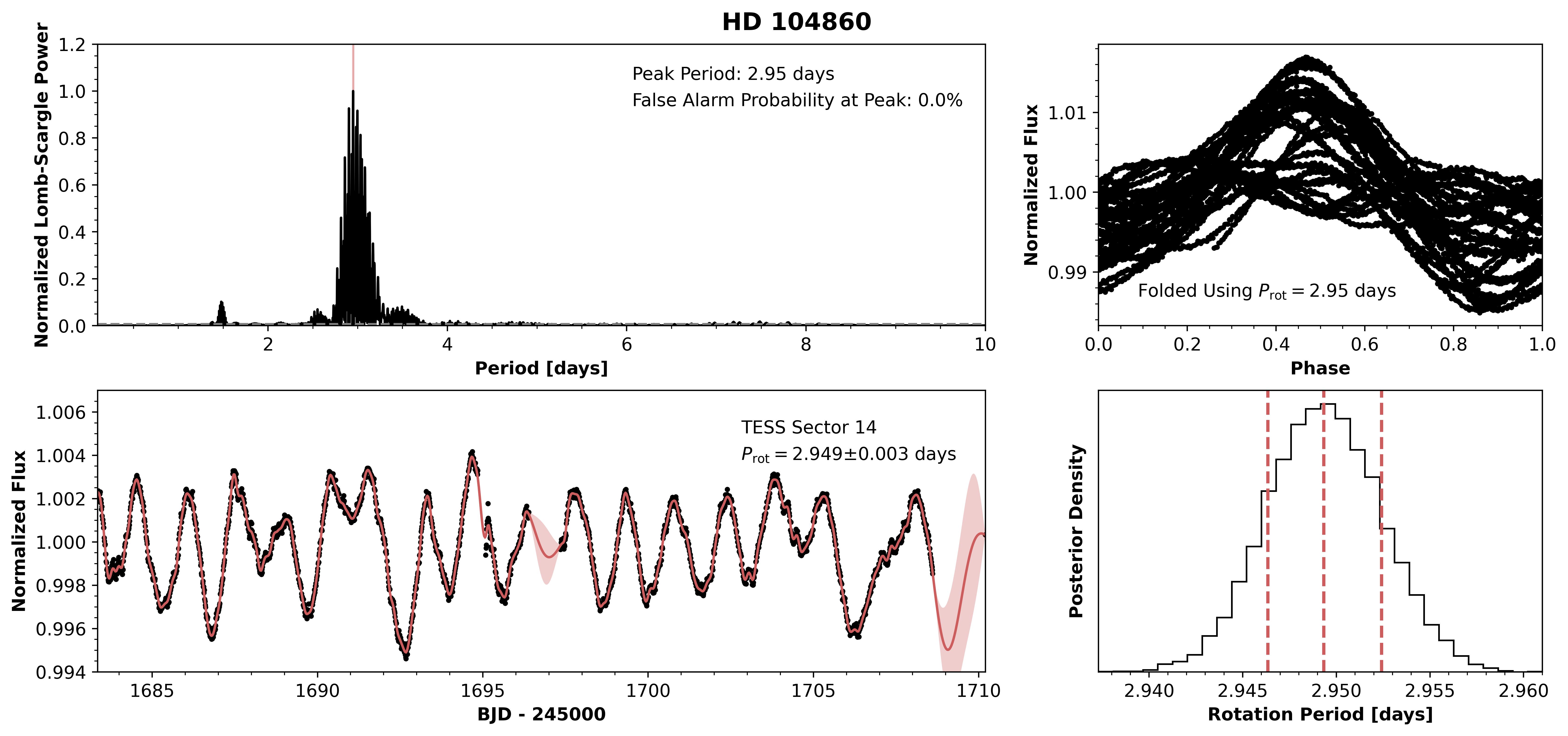}
    \caption{\textit{Top left:} The Lomb-Scargle periodogram for the HD 104860 \textit{TESS} PDCSAP light curve is shown in black. The dashed grey line marks the 1$\%$ FAP level while the shaded red region denotes the $1\sigma$ confidence interval for the rotation period posterior. \textit{Top right:} The phase-folded light curve using the peak period from the Lomb-Scargle periodogram. \textit{Bottom left:} The black points show data from \textit{TESS} Sector 14 while the red line and shaded region mark the mean and $1\sigma$ confidence interval for the GP model. \textit{Bottom right:} The histogram shows the rotation period posterior derived from the GP model while the dash red lines mark the median and $1\sigma$ interval.}
\end{figure}

\begin{figure}[!h]
    \centering
    \includegraphics[width=\linewidth]{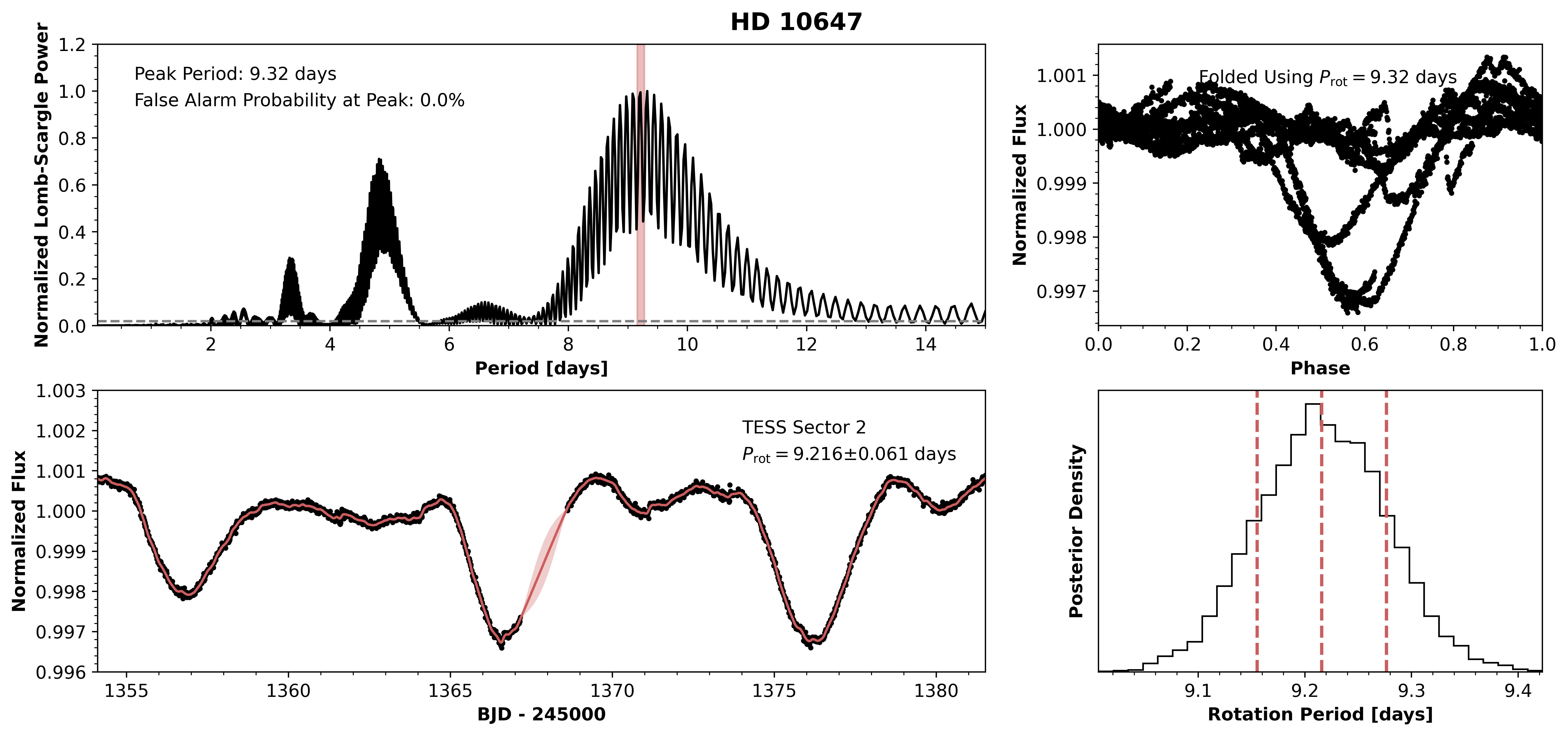}
    \caption{\textit{Top left:} The Lomb-Scargle periodogram for the HD 10647 \textit{TESS} PDCSAP light curve is shown in black. The dashed grey line marks the 1$\%$ FAP level while the shaded red region denotes the $1\sigma$ confidence interval for the rotation period posterior. \textit{Top right:} The phase-folded light curve using the peak period from the Lomb-Scargle periodogram. \textit{Bottom left:} The black points show data from \textit{TESS} Sector 2 while the red line and shaded region mark the mean and $1\sigma$ confidence interval for the GP model. \textit{Bottom right:} The histogram shows the rotation period posterior derived from the GP model while the dash red lines mark the median and $1\sigma$ interval.}
\end{figure}

\begin{figure}[!h]
    \centering
    \includegraphics[width=\linewidth]{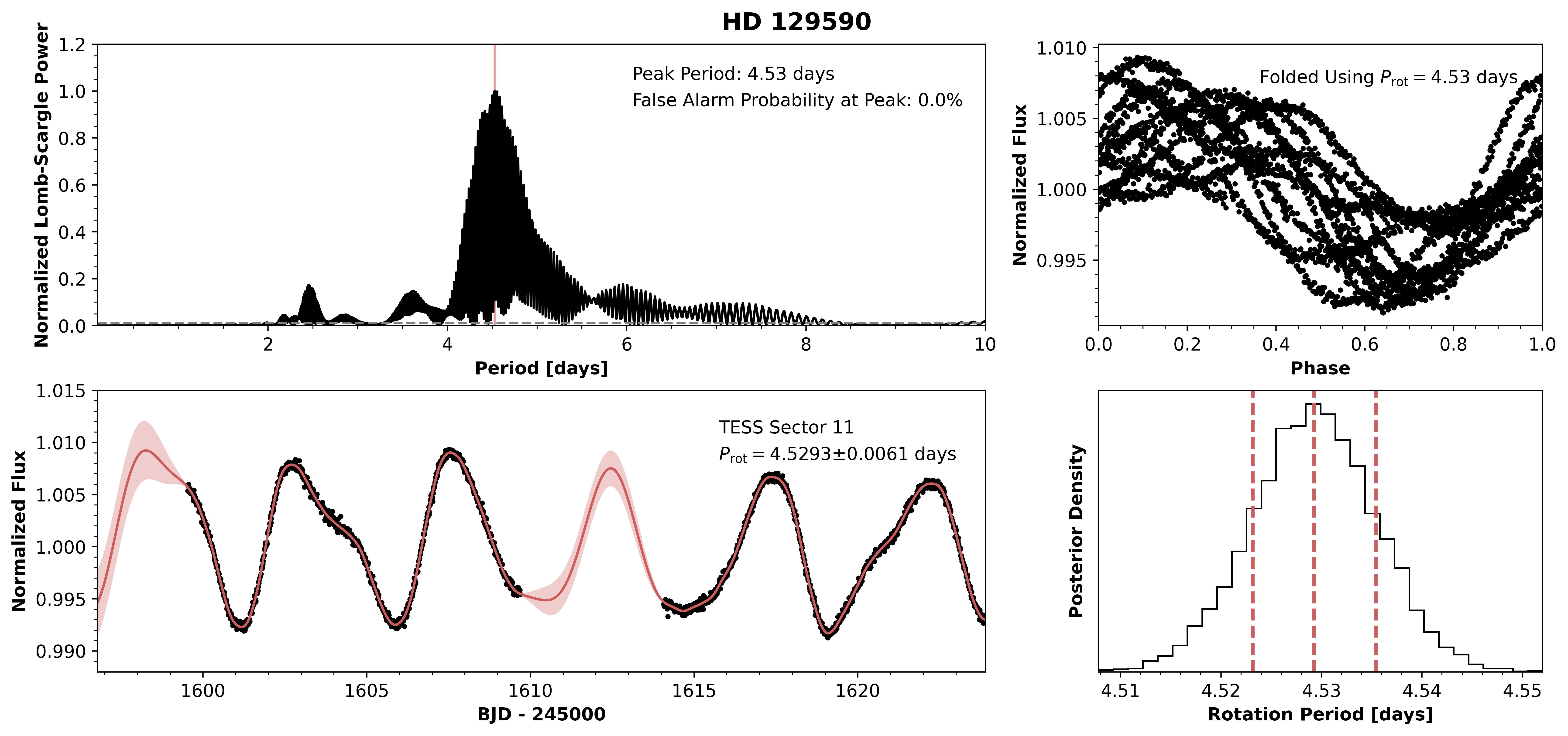}
    \caption{\textit{Top left:} The Lomb-Scargle periodogram for the HD 129590 \textit{TESS} PDCSAP light curve is shown in black. The dashed grey line marks the 1$\%$ FAP level while the shaded red region denotes the $1\sigma$ confidence interval for the rotation period posterior. \textit{Top right:} The phase-folded light curve using the peak period from the Lomb-Scargle periodogram. \textit{Bottom left:} The black points show data from \textit{TESS} Sector 11 while the red line and shaded region mark the mean and $1\sigma$ confidence interval for the GP model. \textit{Bottom right:} The histogram shows the rotation period posterior derived from the GP model while the dash red lines mark the median and $1\sigma$ interval.}
\end{figure}

\begin{figure}[!h]
    \centering
    \includegraphics[width=\linewidth]{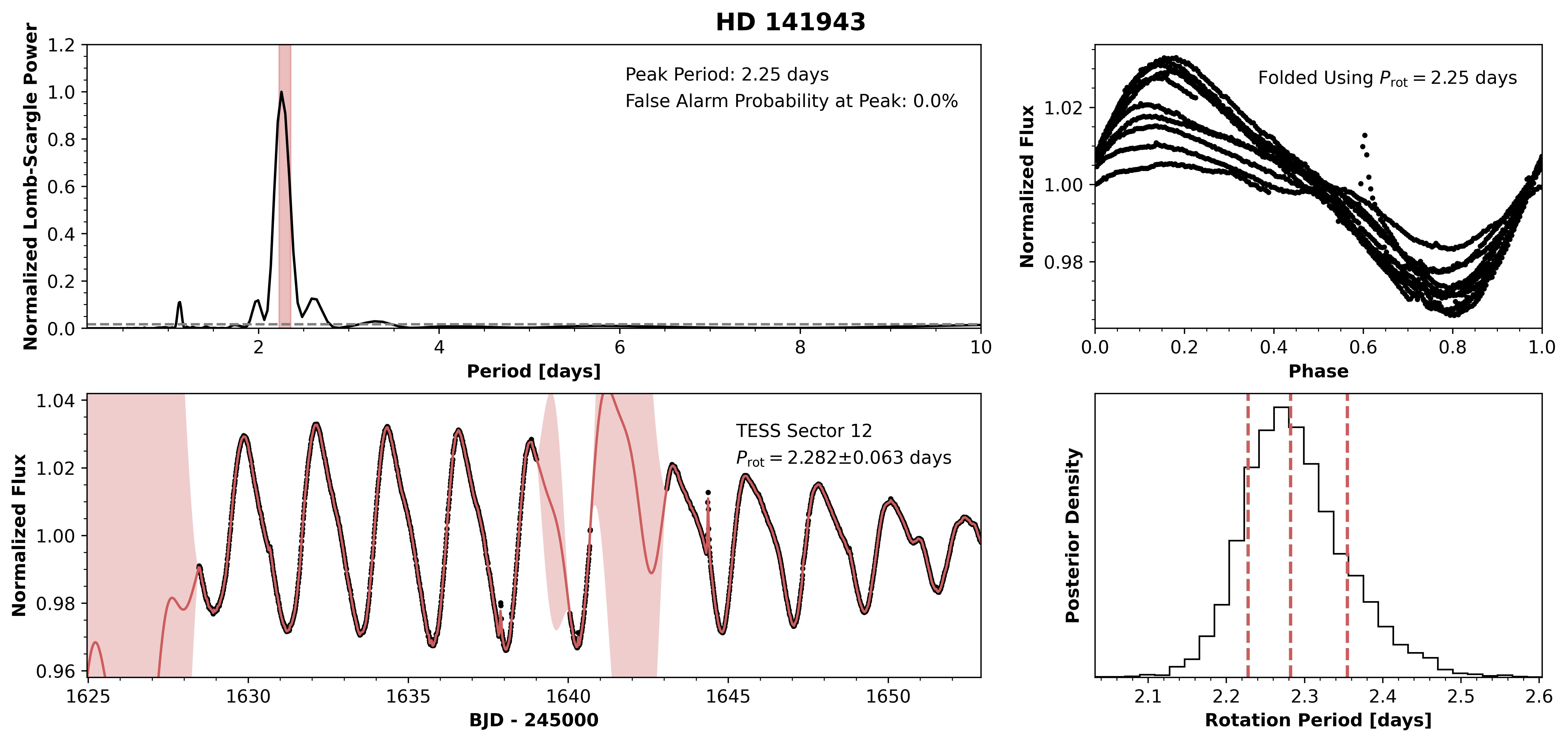}
    \caption{\textit{Top left:} The Lomb-Scargle periodogram for the HD 141943 \textit{TESS} PDCSAP light curve is shown in black. The dashed grey line marks the 1$\%$ FAP level while the shaded red region denotes the $1\sigma$ confidence interval for the rotation period posterior. \textit{Top right:} The phase-folded light curve using the peak period from the Lomb-Scargle periodogram. \textit{Bottom left:} The black points show data from \textit{TESS} Sector 12 while the red line and shaded region mark the mean and $1\sigma$ confidence interval for the GP model. \textit{Bottom right:} The histogram shows the rotation period posterior derived from the GP model while the dash red lines mark the median and $1\sigma$ interval.}
\end{figure}

\begin{figure}[!h]
    \centering
    \includegraphics[width=\linewidth]{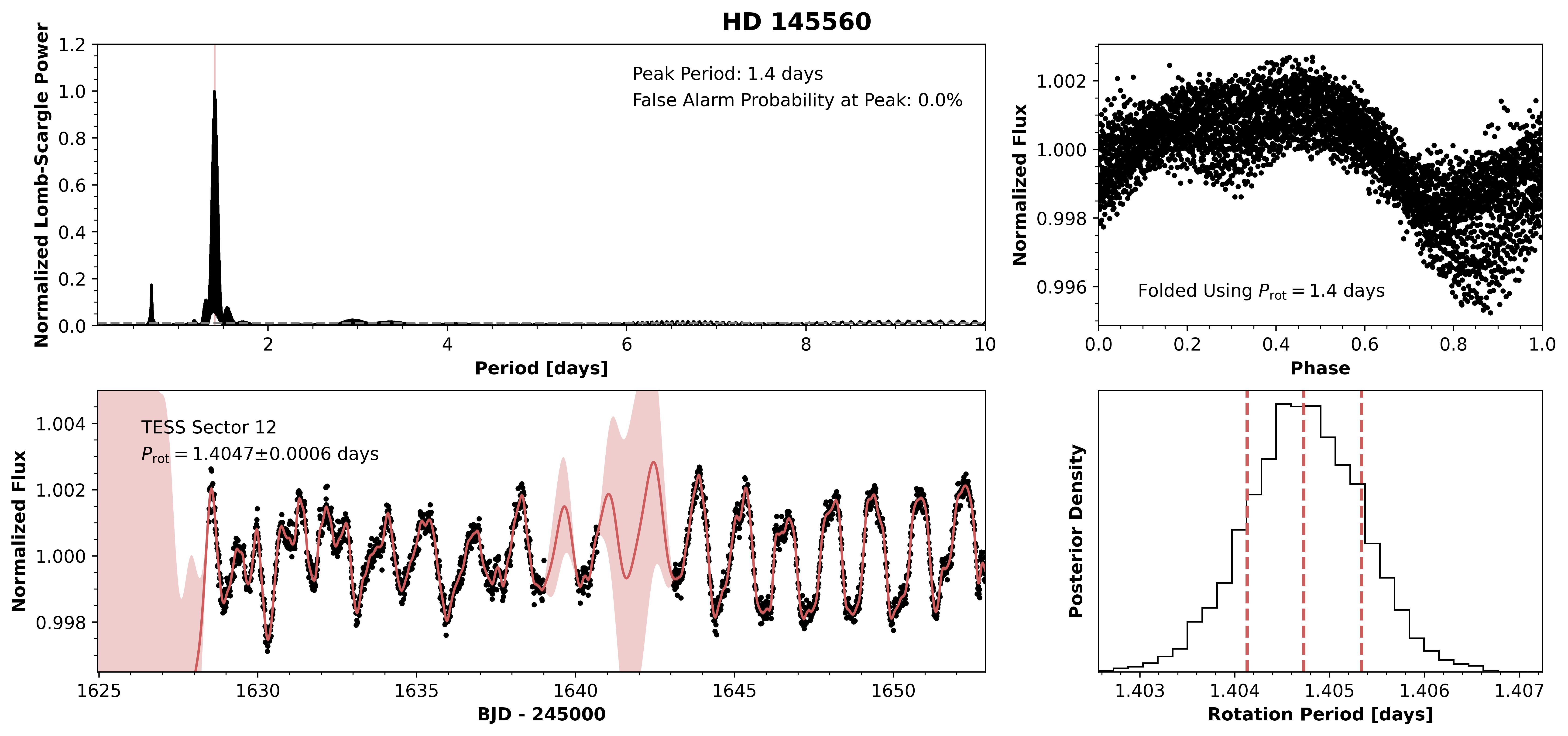}
    \caption{\textit{Top left:} The Lomb-Scargle periodogram for the HD 145560 \textit{TESS} PDCSAP light curve is shown in black. The dashed grey line marks the 1$\%$ FAP level while the shaded red region denotes the $1\sigma$ confidence interval for the rotation period posterior. \textit{Top right:} The phase-folded light curve using the peak period from the Lomb-Scargle periodogram. \textit{Bottom left:} The black points show data from \textit{TESS} Sector 12 while the red line and shaded region mark the mean and $1\sigma$ confidence interval for the GP model. \textit{Bottom right:} The histogram shows the rotation period posterior derived from the GP model while the dash red lines mark the median and $1\sigma$ interval.}
\end{figure}

\begin{figure}[!h]
    \centering
    \includegraphics[width=\linewidth]{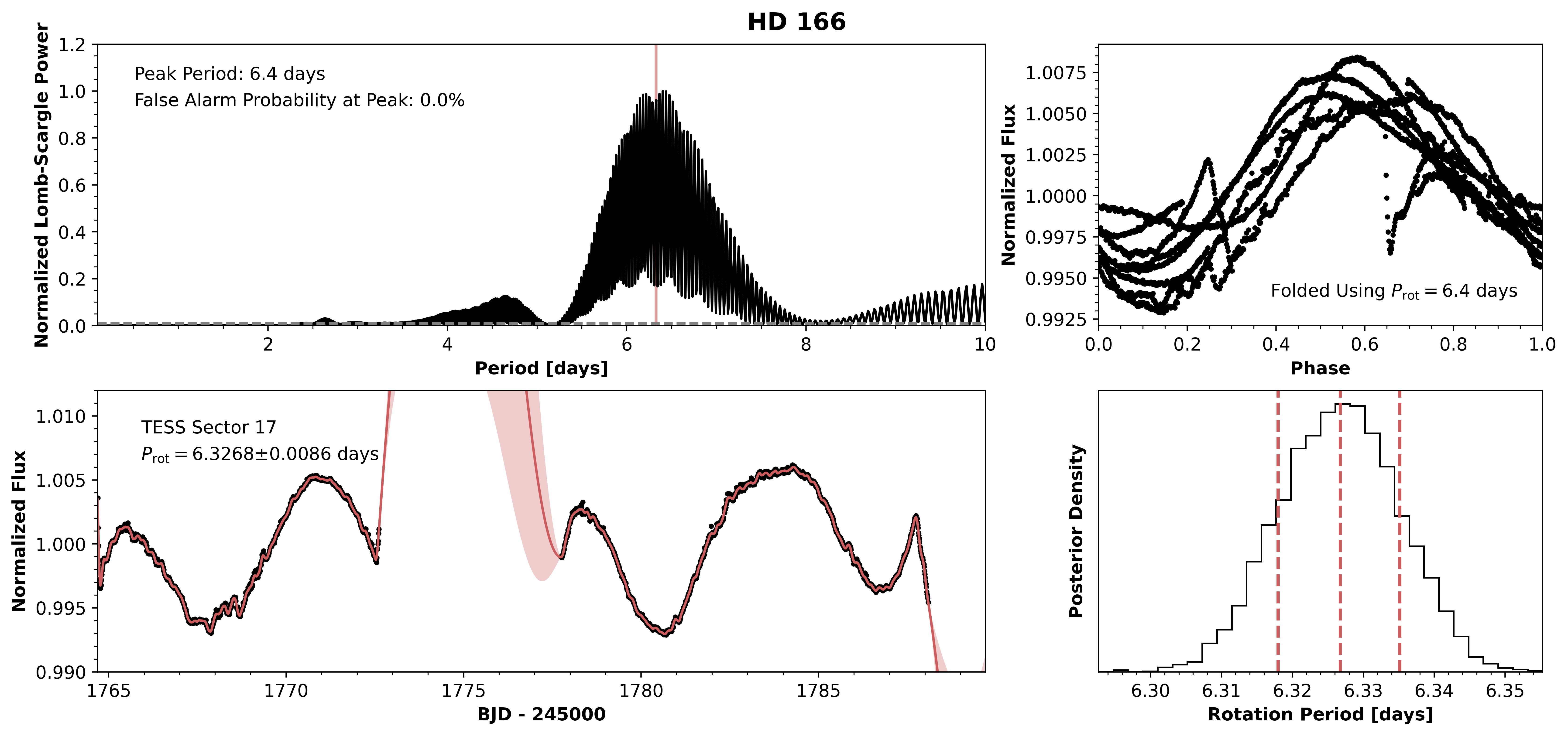}
    \caption{\textit{Top left:} The Lomb-Scargle periodogram for the HD 166 \textit{TESS} PDCSAP light curve is shown in black. The dashed grey line marks the 1$\%$ FAP level while the shaded red region denotes the $1\sigma$ confidence interval for the rotation period posterior. \textit{Top right:} The phase-folded light curve using the peak period from the Lomb-Scargle periodogram. \textit{Bottom left:} The black points show data from \textit{TESS} Sector 17 while the red line and shaded region mark the mean and $1\sigma$ confidence interval for the GP model. \textit{Bottom right:} The histogram shows the rotation period posterior derived from the GP model while the dash red lines mark the median and $1\sigma$ interval.}
\end{figure}

\begin{figure}[!h]
    \centering
    \includegraphics[width=\linewidth]{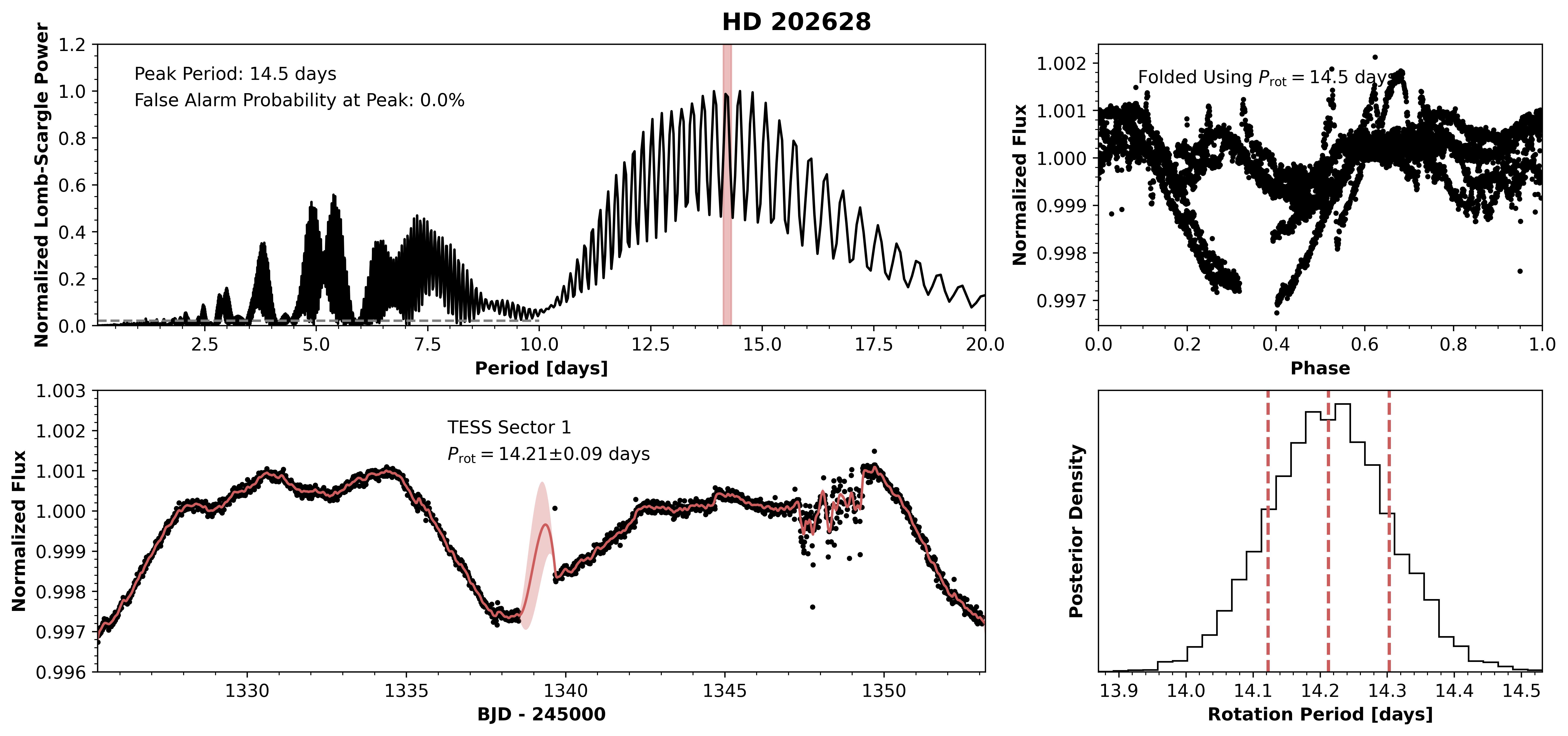}
    \caption{\textit{Top left:} The Lomb-Scargle periodogram for the HD 202628 \textit{TESS} PDCSAP light curve is shown in black. The dashed grey line marks the 1$\%$ FAP level while the shaded red region denotes the $1\sigma$ confidence interval for the rotation period posterior. \textit{Top right:} The phase-folded light curve using the peak period from the Lomb-Scargle periodogram. \textit{Bottom left:} The black points show data from \textit{TESS} Sector 1 while the red line and shaded region mark the mean and $1\sigma$ confidence interval for the GP model. \textit{Bottom right:} The histogram shows the rotation period posterior derived from the GP model while the dash red lines mark the median and $1\sigma$ interval.}
\end{figure}

\begin{figure}[!h]
    \centering
    \includegraphics[width=\linewidth]{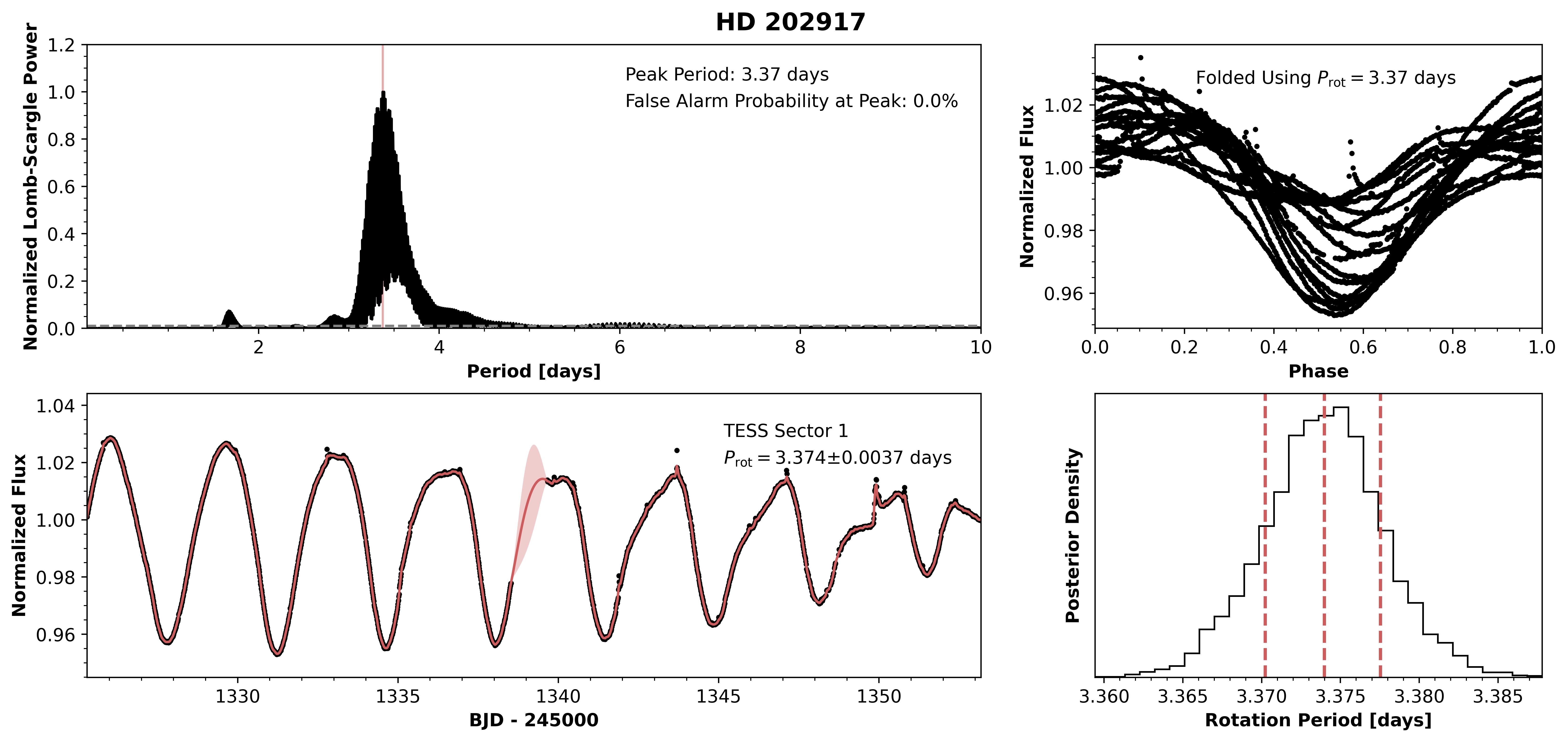}
    \caption{\textit{Top left:} The Lomb-Scargle periodogram for the HD 202917 \textit{TESS} PDCSAP light curve is shown in black. The dashed grey line marks the 1$\%$ FAP level while the shaded red region denotes the $1\sigma$ confidence interval for the rotation period posterior. \textit{Top right:} The phase-folded light curve using the peak period from the Lomb-Scargle periodogram. \textit{Bottom left:} The black points show data from \textit{TESS} Sector 1 while the red line and shaded region mark the mean and $1\sigma$ confidence interval for the GP model. \textit{Bottom right:} The histogram shows the rotation period posterior derived from the GP model while the dash red lines mark the median and $1\sigma$ interval.}
\end{figure}

\begin{figure}[!h]
    \centering
    \includegraphics[width=\linewidth]{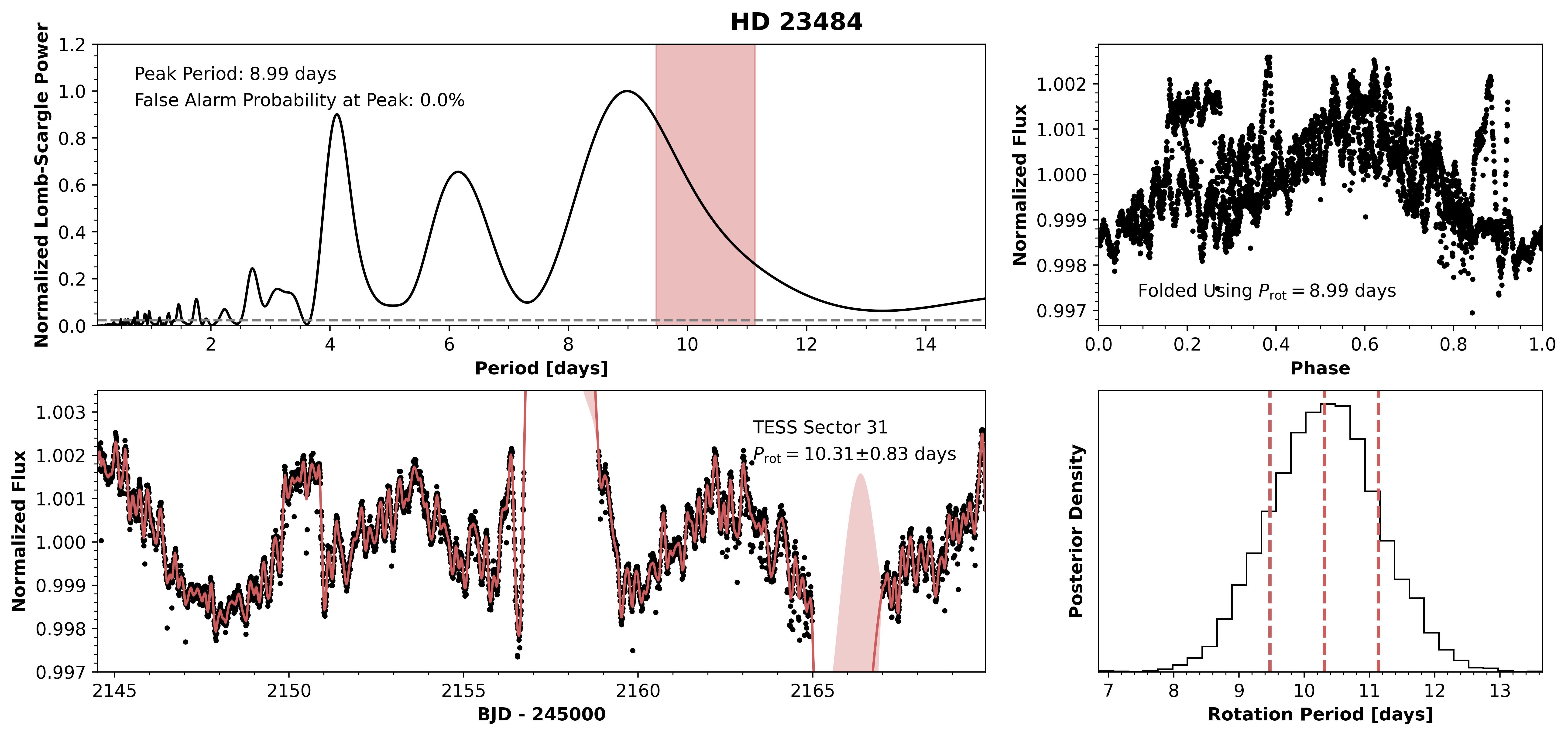}
    \caption{\textit{Top left:} The Lomb-Scargle periodogram for the HD 23484 \textit{TESS} PDCSAP light curve is shown in black. The dashed grey line marks the 1$\%$ FAP level while the shaded red region denotes the $1\sigma$ confidence interval for the rotation period posterior. \textit{Top right:} The phase-folded light curve using the peak period from the Lomb-Scargle periodogram. \textit{Bottom left:} The black points show data from \textit{TESS} Sector 31 while the red line and shaded region mark the mean and $1\sigma$ confidence interval for the GP model. \textit{Bottom right:} The histogram shows the rotation period posterior derived from the GP model while the dash red lines mark the median and $1\sigma$ interval.}
\end{figure}

\begin{figure}[!h]
    \centering
    \includegraphics[width=\linewidth]{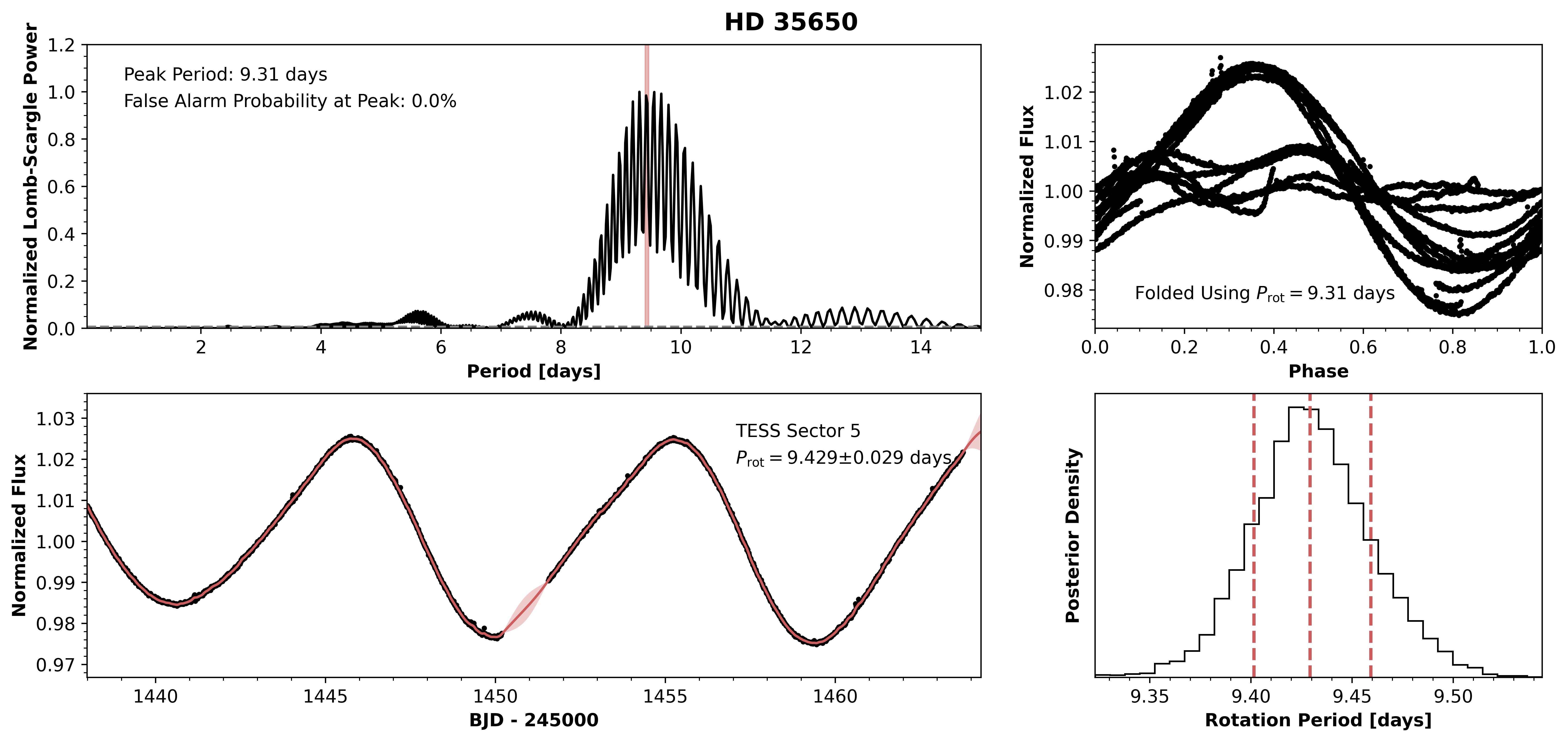}
    \caption{\textit{Top left:} The Lomb-Scargle periodogram for the HD 35650 \textit{TESS} PDCSAP light curve is shown in black. The dashed grey line marks the 1$\%$ FAP level while the shaded red region denotes the $1\sigma$ confidence interval for the rotation period posterior. \textit{Top right:} The phase-folded light curve using the peak period from the Lomb-Scargle periodogram. \textit{Bottom left:} The black points show data from \textit{TESS} Sector 5 while the red line and shaded region mark the mean and $1\sigma$ confidence interval for the GP model. \textit{Bottom right:} The histogram shows the rotation period posterior derived from the GP model while the dash red lines mark the median and $1\sigma$ interval.}
\end{figure}

\begin{figure}[!h]
    \centering
    \includegraphics[width=\linewidth]{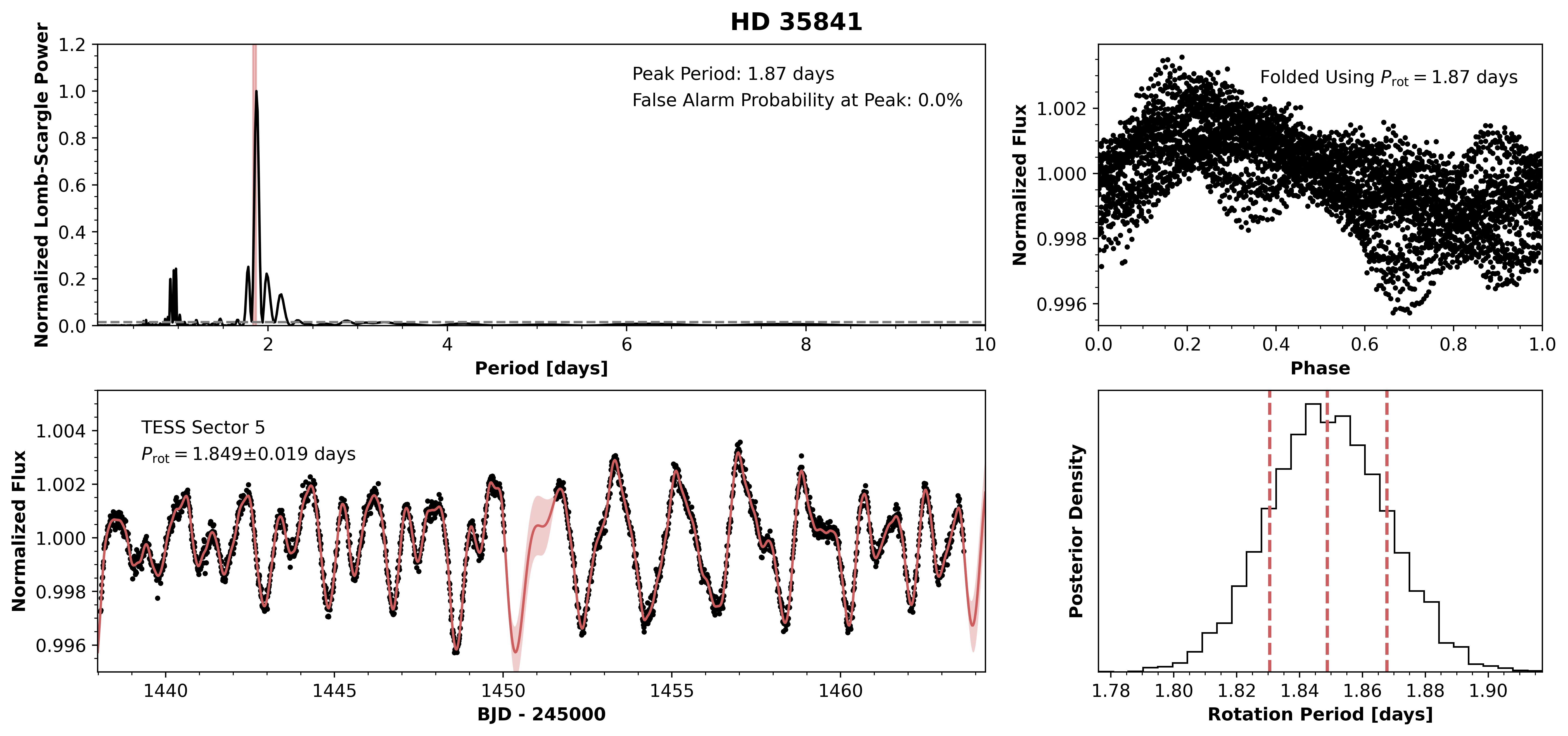}
    \caption{\textit{Top left:} The Lomb-Scargle periodogram for the HD 35841 \textit{TESS} PDCSAP light curve is shown in black. The dashed grey line marks the 1$\%$ FAP level while the shaded red region denotes the $1\sigma$ confidence interval for the rotation period posterior. \textit{Top right:} The phase-folded light curve using the peak period from the Lomb-Scargle periodogram. \textit{Bottom left:} The black points show data from \textit{TESS} Sector 5 while the red line and shaded region mark the mean and $1\sigma$ confidence interval for the GP model. \textit{Bottom right:} The histogram shows the rotation period posterior derived from the GP model while the dash red lines mark the median and $1\sigma$ interval.}
\end{figure}

\begin{figure}[!h]
    \centering
    \includegraphics[width=\linewidth]{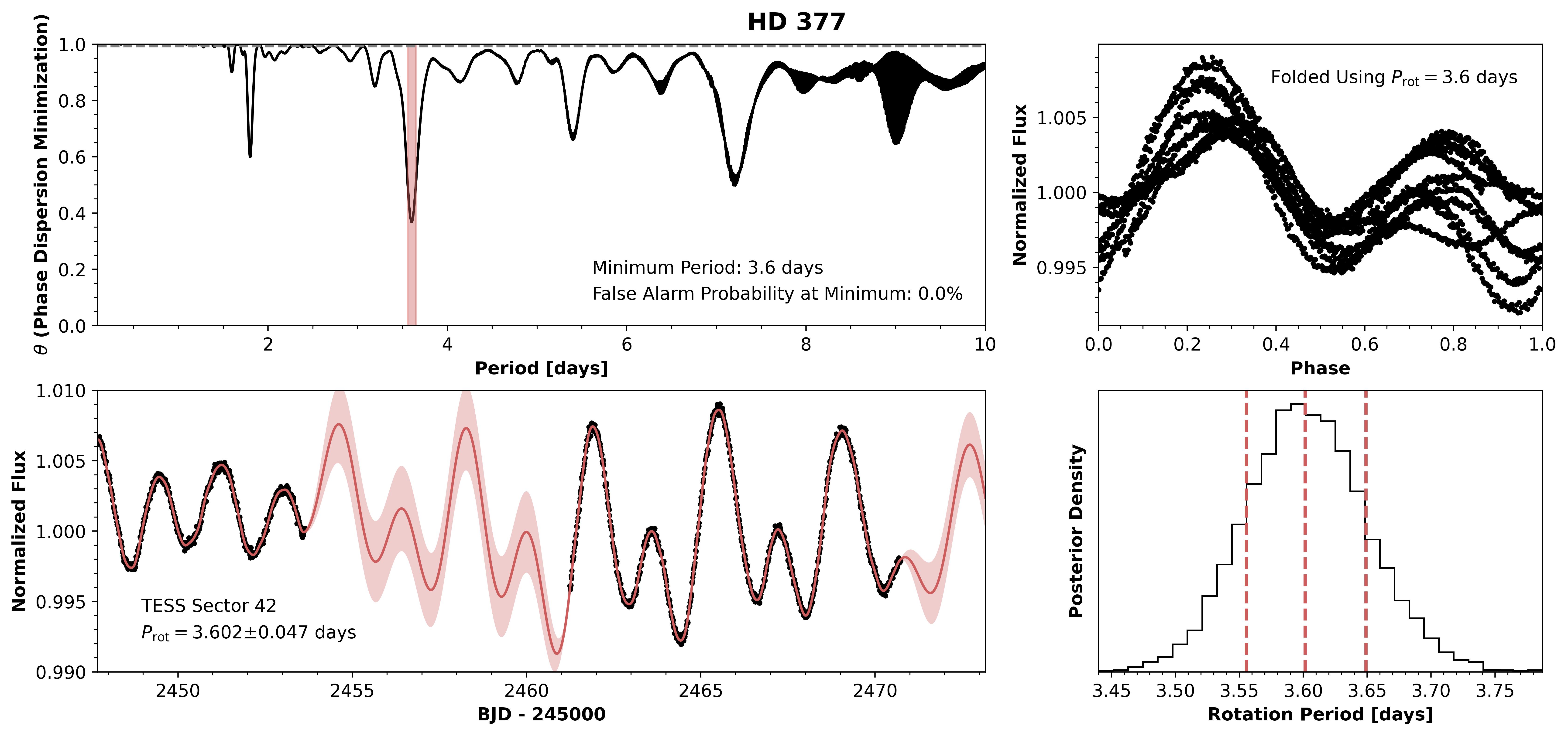}
    \caption{\textit{Top left:} The phase dispersion minimization periodogram for the HD 377 \textit{TESS} PDCSAP light curve is shown in black. The dashed grey line marks the 1$\%$ FAP level while the shaded red region denotes the $1\sigma$ confidence interval for the rotation period posterior. \textit{Top right:} The phase-folded light curve using the peak period from the Lomb-Scargle periodogram. \textit{Bottom left:} The black points show data from \textit{TESS} Sector 5 while the red line and shaded region mark the mean and $1\sigma$ confidence interval for the GP model. \textit{Bottom right:} The histogram shows the rotation period posterior derived from the GP model while the dash red lines mark the median and $1\sigma$ interval.}
\end{figure}

\begin{figure}[!h]
    \centering
    \includegraphics[width=\linewidth]{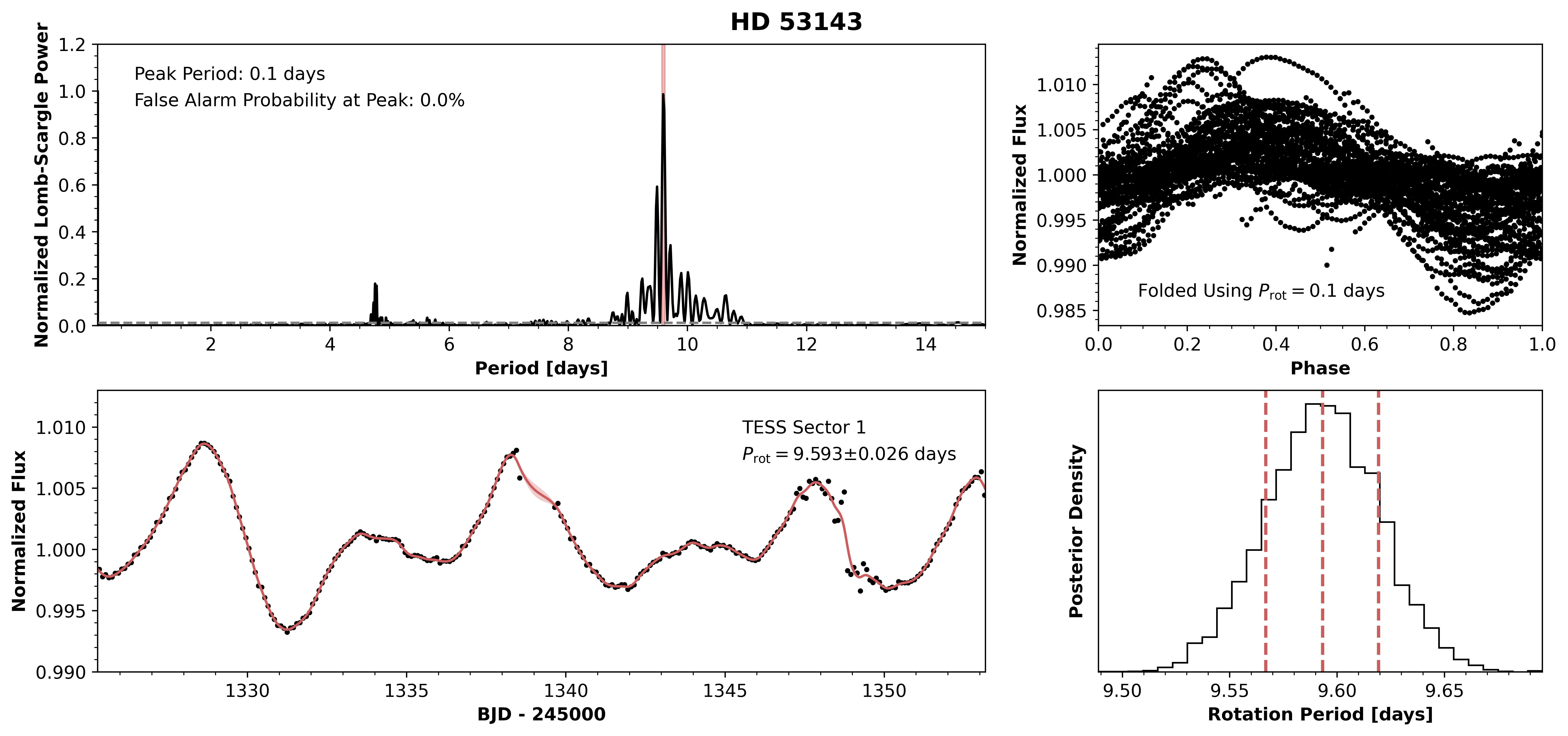}
    \caption{\textit{Top left:} The Lomb-Scargle periodogram for the HD 53143 \textit{TESS} PDCSAP light curve is shown in black. The dashed grey line marks the 1$\%$ FAP level while the shaded red region denotes the $1\sigma$ confidence interval for the rotation period posterior. \textit{Top right:} The phase-folded light curve using the peak period from the Lomb-Scargle periodogram. \textit{Bottom left:} The black points show data from \textit{TESS} Sector 1 while the red line and shaded region mark the mean and $1\sigma$ confidence interval for the GP model. \textit{Bottom right:} The histogram shows the rotation period posterior derived from the GP model while the dash red lines mark the median and $1\sigma$ interval.}
\end{figure}

\begin{figure}[!h]
    \centering
    \includegraphics[width=\linewidth]{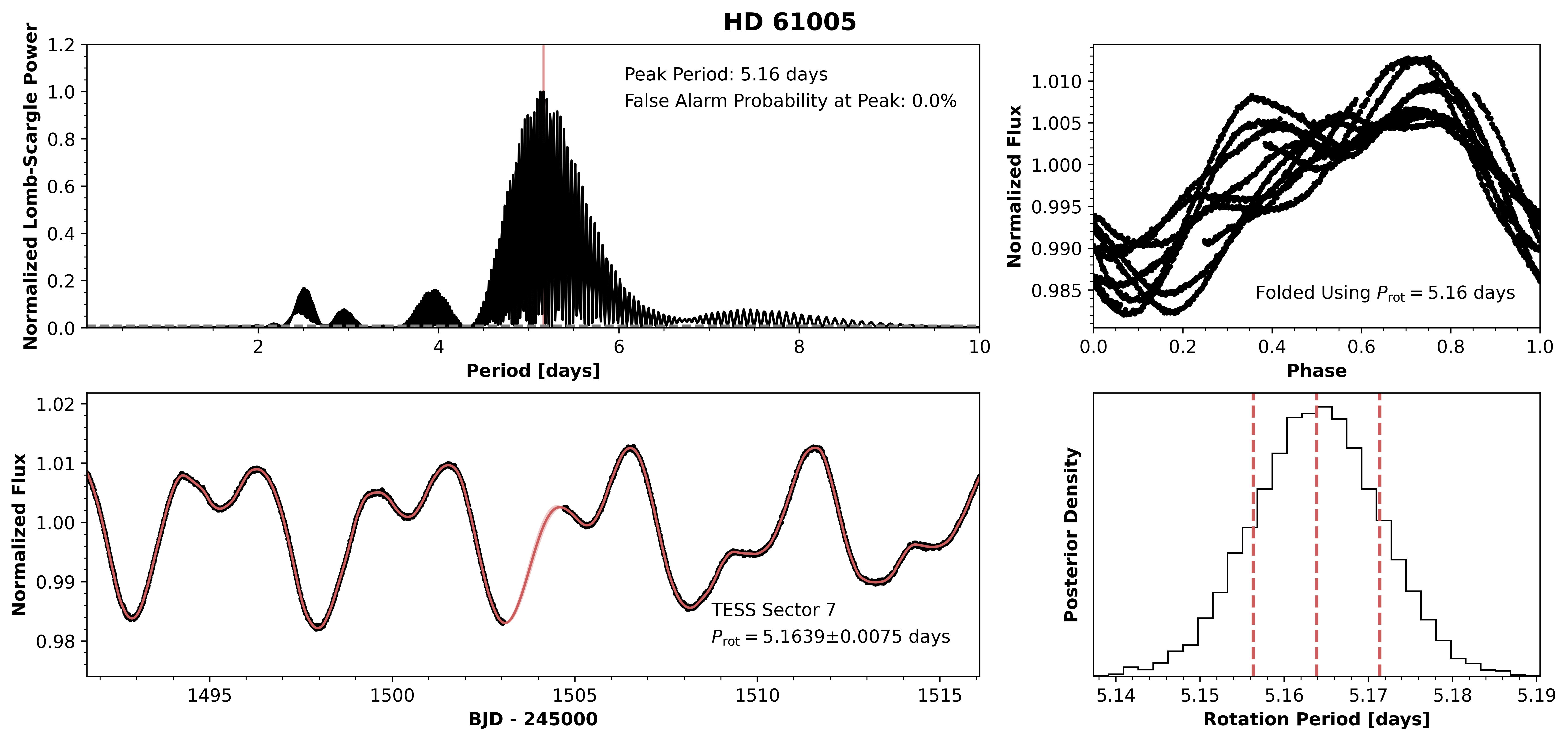}
    \caption{\textit{Top left:} The Lomb-Scargle periodogram for the HD 61005 \textit{TESS} PDCSAP light curve is shown in black. The dashed grey line marks the 1$\%$ FAP level while the shaded red region denotes the $1\sigma$ confidence interval for the rotation period posterior. \textit{Top right:} The phase-folded light curve using the peak period from the Lomb-Scargle periodogram. \textit{Bottom left:} The black points show data from \textit{TESS} Sector 7 while the red line and shaded region mark the mean and $1\sigma$ confidence interval for the GP model. \textit{Bottom right:} The histogram shows the rotation period posterior derived from the GP model while the dash red lines mark the median and $1\sigma$ interval.}
\end{figure}

\begin{figure}[!h]
    \centering
    \includegraphics[width=\linewidth]{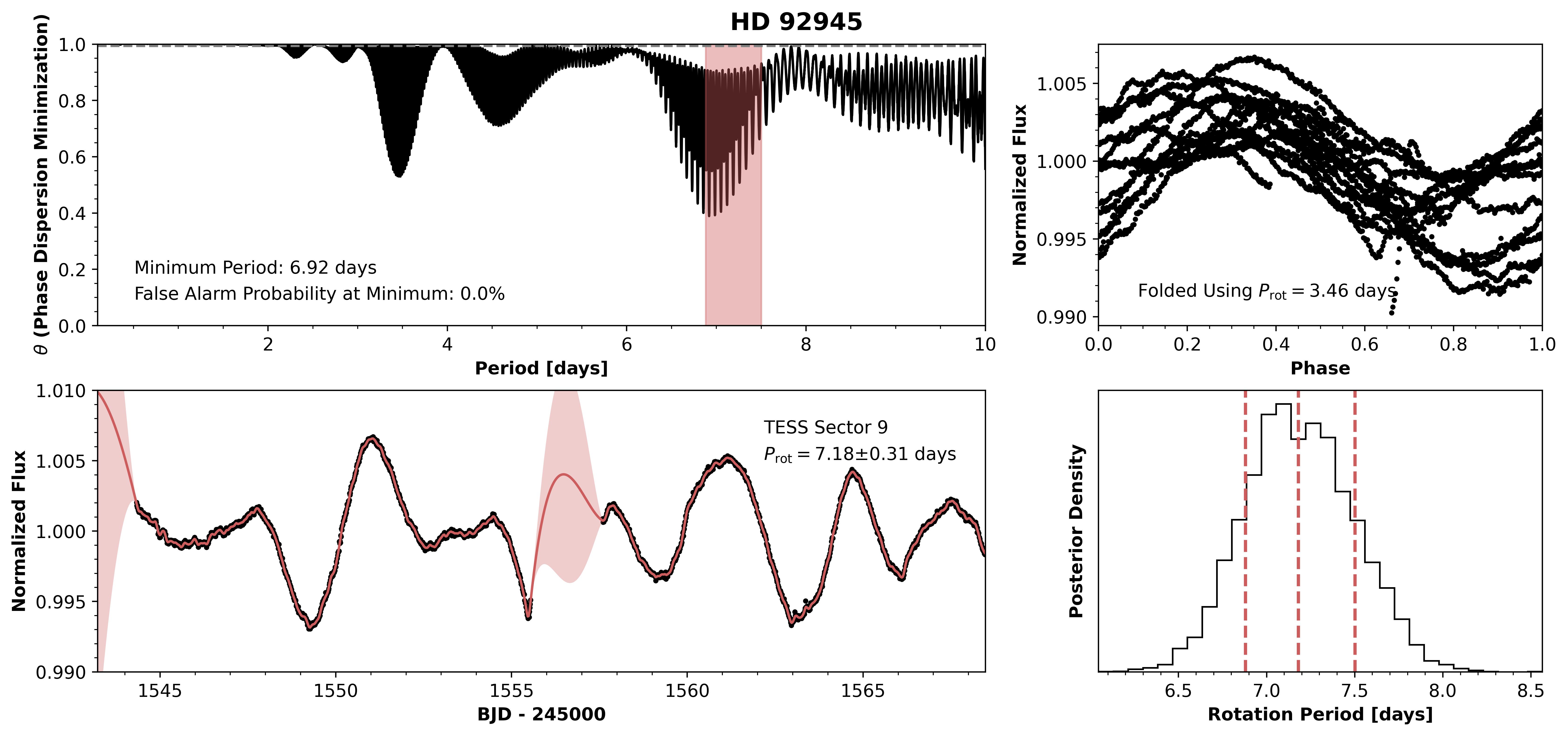}
    \caption{\textit{Top left:} The phase dispersion minimization periodogram for the HD 92945 \textit{TESS} PDCSAP light curve is shown in black. The dashed grey line marks the 1$\%$ FAP level while the shaded red region denotes the $1\sigma$ confidence interval for the rotation period posterior. \textit{Top right:} The phase-folded light curve using the peak period from the Lomb-Scargle periodogram. \textit{Bottom left:} The black points show data from \textit{TESS} Sector 9 while the red line and shaded region mark the mean and $1\sigma$ confidence interval for the GP model. \textit{Bottom right:} The histogram shows the rotation period posterior derived from the GP model while the dash red lines mark the median and $1\sigma$ interval.}
\end{figure}

\begin{figure}[!h]
    \centering
    \includegraphics[width=\linewidth]{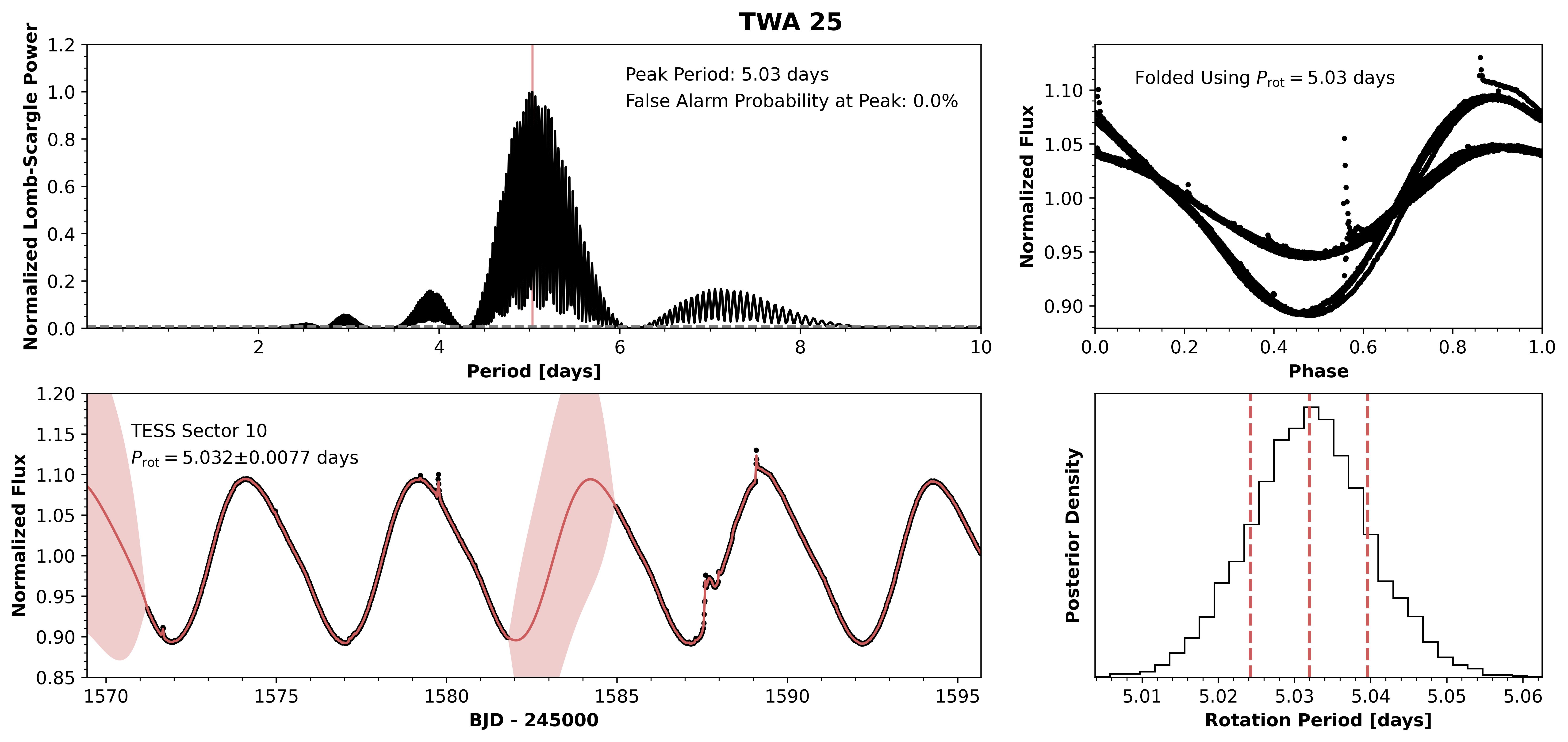}
    \caption{\textit{Top left:} The Lomb-Scargle periodogram for the TWA 25 \textit{TESS} PDCSAP light curve is shown in black. The dashed grey line marks the 1$\%$ FAP level while the shaded red region denotes the $1\sigma$ confidence interval for the rotation period posterior. \textit{Top right:} The phase-folded light curve using the peak period from the Lomb-Scargle periodogram. \textit{Bottom left:} The black points show data from \textit{TESS} Sector 10 while the red line and shaded region mark the mean and $1\sigma$ confidence interval for the GP model. \textit{Bottom right:} The histogram shows the rotation period posterior derived from the GP model while the dash red lines mark the median and $1\sigma$ interval.}
\end{figure}

\clearpage

\bibliography{refs}
\bibliographystyle{aasjournal}

\end{document}